\documentclass[10pt,twocolumn,superscriptaddress,amssymb,amsmath,aps,pra]{revtex4-1}

\usepackage{graphicx}

\begin{document}

\title{Time-dependent theory of optical electro- and magnetostriction}
\date{February 28, 2023}
\author{Mikko Partanen}
\affiliation{Photonics Group, Department of Electronics and Nanoengineering, Aalto University, P.O. Box 13500, 00076 Aalto, Finland}
\author{Bruno Anghinoni}
\affiliation{Department of Physics, Universidade Estadual de Maring\'a, Maring\'a, PR 87020-900, Brazil}
\author{Nelson G. C. Astrath}
\affiliation{Department of Physics, Universidade Estadual de Maring\'a, Maring\'a, PR 87020-900, Brazil}
\author{Jukka Tulkki}
\affiliation{Engineered Nanosystems Group, School of Science, Aalto University, P.O. Box 12200, 00076 Aalto, Finland}

\begin{abstract}
Electrostriction, the deformation of dielectric materials under the influence of an electric field, is of continuous interest in optics. The classic experiment by Hakim and Higham [Proc.~Phys.~Soc.~80, 190 (1962)] for a stationary field supports a different formula of the electrostrictive force density than the recent experiment by Astrath \emph{et al.}~[Light Sci.~Appl.~11, 103 (2022)] for an optical field. In this work, we study the origin of this difference by developing a time-dependent covariant theory of optical force densities in photonic materials. When a light pulse propagates in a bulk dielectric, the field-induced force density consists of two parts: (i) The optical wave momentum force density $\mathbf{f}_\mathrm{owm}$ carries the wave momentum of light and drives forward a mass density wave of the covariant coupled field-material state of light. (ii) The optostrictive force density $\mathbf{f}_\mathrm{ost}$ arises from the atomic density dependence of the electric and magnetic field energy densities. It represents an optical Lorentz-force-law-based generalization of the electro- and magnetostrictive force densities well known for static electromagnetic fields and derived from the principle of virtual work. Since the work done by $\mathbf{f}_\mathrm{ost}$ is not equal to the change of the field energy density during the contraction of the material, we have to describe this difference with optostriction-related dissipation terms to fulfill the energy conservation. The detailed physical model of the dissipation is left for further works. The optostrictive force density can be understood in terms of field-induced pair interactions inside the material. Because of the related action and reaction effects, this force density cannot contribute to the net momentum transfer of the optical field. The theory is used to simulate the propagation of a Gaussian light pulse through a dielectric material. We calculate the electric and magnetic fields of the Gaussian light pulse from Maxwell's equations and simultaneously solve Newton’s equation of motion of atoms to find how the velocity and displacement fields of atoms develop as a function of time under the influence of the field-induced force density.
\end{abstract}

\maketitle

\section{Introduction}

Electrostriction is a phenomenon where materials mechanically deform under the action of an applied electric field in a way that the magnitude of this deformation is proportional to the square of the electric field \cite{Landau1984}. The magnetic analog is called magnetostriction. The theory of electrostriction dates back to the early work of Helmholtz in 1881 \cite{Helmholtz1881}. Despite the long history, the knowledge of electrostriction in optical fields has remained in many ways incomplete due to the challenges in measuring weak optical forces inside materials with an accuracy that would enable detailed comparison with the theory. This situation is, however, changing as new technologies are continuously developed.

An interesting recent discovery by Astrath \emph{et al.} \cite{Astrath2022} has shown that the conventional theory of electrostriction, which explains the results of the classic Hakim-Higham experiment for static fields \cite{Hakim1962}, does not reproduce the magnitude of the electrostriction effect in water at optical frequencies. The difference between the conventional theory of electro- and magnetostriction for static fields and the corresponding theory for time-dependent optical fields and their relation to recent experiments is the starting point of the present work. Several experiments of electrostriction at optical frequencies have been carried out also previously \cite{Melloni1998,Fellegara1997,Fellegara1998,Buckland1997,Buckland1997,Godbout2000,Townsend1996,Dianov1990,Hui2015,Cai2013,Beugnot2012}, but their accuracy has not enabled detailed analysis of the time-dependence of the optical electro- and magnetostriction effects. We also point out that most theoretical and experimental studies of the static electro- and magnetostriction correspond to the thermodynamical equilibrium limit. Experimental studies of electrostriction have also revealed interesting phenomena, such as the giant electrostriction \cite{Zhang2022,ChenBo2018,Wang2022,Yu2022,ChenXi2022,Zhang2021,Wu2020,Huang2019,Jin2019,Karvounis2018,Zhang1998}, negative electrostriction \cite{Yamashita2018,Neumayer2019}, and deformations of liquid crystals \cite{Guo2020,ZhangY2022} and biological cells \cite{Torbati2022}. On the theoretical side, electrostriction has also been under extensive study \cite{Wang2016a,Jimenez2011,Khakpour2021,Lee2005,Rakich2010,Rakich2011,Gojani2014,Smith2015,Hashimoto2018,Pennec2014,Tanner2021,Lai1981,Torchigin2019,Suo2010,WangJ2010,Washimi1976,Zhao2008,Li2014}.

When a light pulse propagates in a bulk dielectric, the field-induced force density consists of two parts: (i) The optical wave momentum force density $\mathbf{f}_\mathrm{owm}$ carries the wave momentum of light and drives forward a mass density wave of the covariant coupled field-material state of light \cite{Partanen2022b,Partanen2017c,Partanen2021b,Partanen2019a,Partanen2019b}. It has been extensively studied regarding its relation to the Abraham-Minkowski controversy of the momentum of light \cite{Leonhardt2006a,Pfeifer2007,Barnett2010b,Kemp2011,Milonni2010,Bliokh2017a,Burt1973,Mansuripur2010,Barnett2010a,Leonhardt2014,Bliokh2017b}. (ii) The optostrictive force density $\mathbf{f}_\mathrm{ost}$ arises from the atomic density dependence of the electric and magnetic field energy densities. The starting point of the present work is the recently developed mass-polariton theory of light \cite{Partanen2022b,Partanen2017c,Partanen2021b,Partanen2019a,Partanen2019b}, which enables a relativistically consistent study of $\mathbf{f}_\mathrm{owm}$ and the momentum transfer associated with the propagation of light. We generalize the mass-polariton theory to include $\mathbf{f}_\mathrm{ost}$ in a way that preserves the relativistic covariance when both force densities are included in the theory.

The force density $\mathbf{f}_\mathrm{ost}$ has conventionally been investigated in the limit of thermodynamics and quasi-static fields, and it has been extended to time-averaged optical fields \cite{Boyd2008}. As briefly discussed in Ref.~\cite{Partanen2022b}, the full time- and position-dependent relativistically consistent theory of $\mathbf{f}_\mathrm{ost}$, in the optical regime, has not been presented in detail previously to the best of our knowledge. The present theory enables a formulation of the optical electrostriction and optical magnetostriction in terms of time- and space-dependent force fields without the need to introduce time or harmonic averages. The unified theory of the time-dependent force density of the optical field on the material has obvious potential in tailoring acousto-optical coupling \cite{Wissmeyer2018,Manohar2016,Tam1983,Lai1982,Patel1981,Davies1993,Pozar2018,Pozar2013} for both fundamental science and technological needs.

The Lorentz-force-law-based derivation of the optical force density leads to an expression of the sum of $\mathbf{f}_\mathrm{owm}$ and $\mathbf{f}_\mathrm{ost}$. The origin of $\mathbf{f}_\mathrm{owm}$ relies in the definition of the Poynting vector, describing the optical energy flux, and the covariance principle of light in dielectric and magnetic media \cite{Partanen2022b,Partanen2017c,Partanen2021b,Partanen2019a,Partanen2019b}. In contradistinction, $\mathbf{f}_\mathrm{ost}$ originates from the atomic density dependence of the electric and magnetic field energy densities. Thus, $\mathbf{f}_\mathrm{ost}$ can be traced back to pair interactions, and due to this force and counterforce nature, the resulting momentum densities always integrate to zero over the volume of the material. Therefore, there is no net momentum transfer related to $\mathbf{f}_\mathrm{ost}$.

We assume that the material is lossless, linear, isotropic, optically nondispersive, and that the Clausius-Mossotti relation is satisfied. These conditions exclude complex molecular materials, in which, e.g., giant electrostriction \cite{Zhang2022,ChenBo2018,Wang2022,Yu2022,ChenXi2022,Zhang2021,Wu2020,Huang2019,Jin2019,Karvounis2018,Zhang1998} and negative electrostriction \cite{Yamashita2018,Neumayer2019} have been discovered. Thermophotonic effects are also excluded. Since the investigations of the present work do not deal with thermodynamical equilibrium, we do not consider extensions of the Clausius-Mossotti relation, which include wavelength, temperature, or density-dependent corrections \cite{Harvey1998,Schiebener1990}. Even with the restrictions above, the theory has a wide range of applications in photonics technologies. It is, however, obvious that the theory can be extended to more complex materials, but these extensions are not discussed in detail in the present work.

This work is organized as follows. Section \ref{sec:conservation} presents the conservation laws of the material, fields, and interactions, and how the force density appears in them. Section \ref{sec:fields} summarizes the relations of the macroscopic and local electric and magnetic fields and flux densities and the relations of the polarization and magnetization fields. These relations determine the density dependence of the permittivity and permeability of the material. Section \ref{sec:forcedensity} presents in detail the different contributions of the force density which the material experiences under the influence of an electromagnetic field. Section \ref{sec:simulations} describes simulations of optical electrostriction in silicon. Section \ref{sec:tensors} presents the stress-energy-momentum (SEM) tensors of different parts of the field-material system and its interactions. Section \ref{sec:discussion} discusses implications of the present theory to the previous mass-polariton theory of light. Finally, conclusions are drawn in Sec.~\ref{sec:conclusions}.

\section{\label{sec:conservation}Conservation laws}

As the foundation of the theory, we use the fundamental conservation laws of energy and momentum. Together with Maxwell's equations and the Lorenz force law, this is a widely used and evidently the most fundamental starting point for describing optical forces \cite{Penfield1967,Partanen2021b,Partanen2022b,Anghinoni2022,Sheppard2014,Kemp2017}. In particular, we focus on writing a relativistically consistent theory, which can be applied to all inertial observers independent of their velocities with respect to the material. This condition imposes strong limitations for possible forms of the SEM tensors and the related force fields as discussed further in Sec.~\ref{sec:tensors}. Accordingly, when writing the conservation laws below and the general expressions of the SEM tensors in Sec.~\ref{sec:tensors}, we assume a general inertial frame. In the presentation of the force density in Sec~\ref{sec:forcedensity}, we present the equations for the laboratory frame, which is a special inertial frame, where the material atoms are at rest before the force density of the electromagnetic field starts to accelerate them. In the presentation below, we split the field-material system into two parts: (1) the material subsystem and (2) the field+interaction subsystem and write the conservation laws for both of these subsystems satisfying the law of action and reaction.

\subsection{Force density and conservation laws of the material}

We assume that the material is composed of identical atoms. For a single atom, the rest mass is denoted by $m_0$, the velocity is $\mathbf{v}_\mathrm{a}$, and the momentum is $\mathbf{p}_\mathrm{a}=\gamma_{\mathbf{v}_\mathrm{a}}m_0\mathbf{v}_\mathrm{a}$, where $\gamma_{\mathbf{v}_\mathrm{a}}=1/\sqrt{1-|\mathbf{v}_\mathrm{a}|^2/c^2}$ is the Lorentz factor, in which $c$ is the speed of light in vacuum. In the laboratory frame, the atomic velocities caused by the optical field are very small, and thus the Lorentz factor is close to unity. However, for the consideration of relativistic invariance properties later in this work, we use the general form that applies to all inertial observers. The atomic number density is denoted by $n_\mathrm{a}$. The fundamental definition of the force density $\mathbf{f}=n_\mathrm{a}\mathbf{F}$, where $\mathbf{F}$ is the force on a single atom, is given by Newton's equation of motion, written in any inertial frame as \cite{Penfield1967,Partanen2021b}
\begin{equation}
 n_\mathrm{a}\frac{d\mathbf{p}_\mathrm{a}}{dt}=\mathbf{f}.
 \label{eq:Newton}
\end{equation}
First, by using the material derivative $\frac{d}{dt}=\frac{\partial}{\partial t}+\mathbf{v}_\mathrm{a}\cdot\boldsymbol{\nabla}$ to Eq.~\eqref{eq:Newton}, second, multiplying the conservation law of the atomic number density, $\frac{\partial}{\partial t}n_\mathrm{a}+\boldsymbol{\nabla}\cdot(n_\mathrm{a}\mathbf{v}_\mathrm{a})=0$, by $\mathbf{p}_\mathrm{a}$, and, third, adding the resulting two equations side by side gives after some vector algebra \cite{Penfield1967,Partanen2021b}
\begin{equation}
 \frac{\partial\mathbf{G}_\mathrm{mat}}{\partial t}+\boldsymbol{\nabla}\cdot\boldsymbol{\mathcal{T}}_\mathrm{mat}=\mathbf{f}.
 \label{eq:momentumconservation}
\end{equation}
Here $\mathbf{G}_\mathrm{mat}=n_\mathrm{a}\mathbf{p}_\mathrm{a}$ is the momentum density of the material and $\boldsymbol{\mathcal{T}}_\mathrm{mat}=n_\mathrm{a}\mathbf{p}_\mathrm{a}\otimes\mathbf{v}_\mathrm{a}$, where $\otimes$ denotes the outer product of vectors, is the stress tensor of the material, which does not account for the stresses associated to the mechanical or electromagnetically induced pressures. Correspondingly, these phenomena are included through the force density on the right-hand side of Eq.~\eqref{eq:momentumconservation} as described below. Equation \eqref{eq:momentumconservation} is the conservation law of momentum. The conservation law of energy of the material reads
\begin{equation}
 \frac{1}{c^2}\frac{\partial W_\mathrm{mat}}{\partial t}+\boldsymbol{\nabla}\cdot\mathbf{G}_\mathrm{mat}=\frac{\phi}{c^2},
 \label{eq:energyconservation}
\end{equation}
where $W_\mathrm{mat}=\gamma_{\mathbf{v}_\mathrm{a}}m_0c^2$ is the energy density of the material and $\phi=\mathbf{v}_\mathrm{a}\cdot\mathbf{f}$ is the power conversion density of the kinetic energy of the material, in accordance with Eq.~\eqref{eq:momentumconservation}. In previous literature, the material subsystem described above is also called the kinetic subsystem \cite{Penfield1967}.

\subsection{\label{sec:fieldandpressureeffects}Conservation laws of the field and electromagnetic and mechanical pressure effects}

After defining the energy and momentum densities and the stress tensor of the material and their conservation laws in Eqs.~\eqref{eq:momentumconservation} and \eqref{eq:energyconservation}, we will now define the energy density $W_\mathrm{emi}$, the momentum density $\mathbf{G}_\mathrm{emi}$, and the stress tensor $\boldsymbol{\mathcal{T}}_\mathrm{emi}$ and their conservation laws for \emph{the rest of the field-material system}. We call this subsystem the field+interaction subsystem in accordance with previous literature, where it is the sum of the field and interaction subsystems \cite{Penfield1967}. The conservation laws of the momentum and energy of the field+interaction subsystem are written as \cite{Penfield1967,Jackson1999,Landau1984}
\begin{equation}
 \frac{\partial\mathbf{G}_\mathrm{emi}}{\partial t}+\boldsymbol{\nabla}\cdot\boldsymbol{\mathcal{T}}_\mathrm{emi}=-\mathbf{f}.
 \label{eq:conservationf}
\end{equation}
\begin{equation}
 \frac{1}{c^2}\frac{\partial W_\mathrm{emi}}{\partial t}+\boldsymbol{\nabla}\cdot\mathbf{G}_\mathrm{emi}=-\frac{\phi}{c^2},
 \label{eq:conservationphi}
\end{equation}
The force density and power-conversion density are generally nonzero, which means that the field+interaction subsystem is exchanging energy and momentum with the material subsystem, whose conservation laws are given in Eqs.~\eqref{eq:momentumconservation} and \eqref{eq:energyconservation}. Due to the law of action and reaction, there are opposite signs in the momentum conservation laws in Eqs.~\eqref{eq:momentumconservation} and \eqref{eq:conservationf} and the energy conservation laws in Eqs.~\eqref{eq:energyconservation} and \eqref{eq:conservationphi}.

\section{\label{sec:fields}Microscopic and macroscopic fields and the density dependence of permittivity and permeability}

Before studying the electromagnetic force densities in detail, in this section, we consider the dependence of the permittivity and permeability of the material on the atomic density. This dependence originates from the microscopic expressions of the polarization field $\mathbf{P}$ and the magnetization field $\mathbf{M}$ and from their relation to the macroscopic fields \cite{Panofsky1962,Jackson1999}. The macroscopic electric field $\mathbf{E}$ and magnetic field $\mathbf{H}$ are related to the electric flux density $\mathbf{D}$, magnetic flux density $\mathbf{B}$, and the fields $\mathbf{P}$ and $\mathbf{M}$ by the well-known constitutive relations of a nondispersive material, given by
\begin{equation}
 \mathbf{D}=\varepsilon_0\varepsilon_\mathrm{r}\mathbf{E},
 \hspace{0.5cm}\mathbf{P}=\varepsilon_0(\varepsilon_\mathrm{r}-1)\mathbf{E}
 \label{eq:D}
\end{equation}
\begin{equation}
 \mathbf{B}=\mu_0\mu_\mathrm{r}\mathbf{H},
 \hspace{0.5cm}\mathbf{M}=(\mu_\mathrm{r}-1)\mathbf{H}.
 \label{eq:B}
\end{equation}
Here $\varepsilon_0$ is the permittivity of vacuum, $\varepsilon_\mathrm{r}$ is the relative permittivity of the material, $\mu_0$ is the permeability of vacuum, and $\mu_\mathrm{r}$ is the relative permeability of the material. In terms of the relative permittivity and permeability, the refractive index of the material is given by $n=\sqrt{\varepsilon_\mathrm{r}\mu_\mathrm{r}}$.

As conventional, we can write the microscopic expressions of the polarization and magnetization fields as dipole moment densities, given by \cite{Panofsky1962,Jackson1999}
\begin{equation}
 \mathbf{P}=n_\mathrm{a}\mathbf{p}=n_\mathrm{a}\alpha_\mathrm{e}\mathbf{E}_\mathrm{eff},
 \label{eq:P}
\end{equation}
\begin{equation}
 \mathbf{M}=n_\mathrm{a}\mathbf{m}=n_\mathrm{a}\alpha_\mathrm{m}\mathbf{B}_\mathrm{eff}.
 \label{eq:M}
\end{equation}
Here $n_\mathrm{a}$ is the number density of atoms, $\mathbf{p}=\alpha_\mathrm{e}\mathbf{E}_\mathrm{eff}$ is the atomic electric dipole moment, $\mathbf{m}=\alpha_\mathrm{m}\mathbf{B}_\mathrm{eff}$ is the atomic magnetic dipole moment, $\alpha_\mathrm{e}$ and $\alpha_\mathrm{m}$ are the atomic polarizability and magnetizability, which are independent of the atomic density, and $\mathbf{E}_\mathrm{eff}$ and $\mathbf{B}_\mathrm{eff}$ are the effective local electric and magnetic fields at the site of the atom, given for isotropic cubic materials by \cite{Lorentz1952,Panofsky1962,Jackson1999,Griffiths1998,Aspnes1982,Shevchenko2010,Shevchenko2011,Maki1991}
\begin{equation}
 \mathbf{E}_\mathrm{eff}=\mathbf{E}+\frac{\mathbf{P}}{3\varepsilon_0}
 =\mathbf{E}\sum_{l=0}^\infty\Big(\frac{n_\mathrm{a}\alpha_\mathrm{e}}{3\varepsilon_0}\Big)^l\!
 =\frac{\mathbf{E}}{1-\frac{n_\mathrm{a}\alpha_\mathrm{e}}{3\varepsilon_0}},
 \label{eq:Eeff}
\end{equation}
\begin{equation}
 \mathbf{B}_\mathrm{eff}=\mathbf{B}-\frac{2\mu_0\mathbf{M}}{3}
 =\mathbf{B}\sum_{l=0}^\infty\!\Big(\!-\frac{2\mu_0n_\mathrm{a}\alpha_\mathrm{m}}{3}\Big)^l\!\!
 =\frac{\mathbf{B}}{1+\frac{2\mu_0n_\mathrm{a}\alpha_\mathrm{m}}{3}}.
 \label{eq:Beff}
\end{equation}
The calculation of the local fields for other symmetries of the materials, e.g., for anisotropic materials, is discussed in Ref.~\cite{Aspnes1982}.

Using the relation between $\mathbf{P}$ and $\mathbf{E}$ in Eq.~\eqref{eq:D} together with Eqs.~\eqref{eq:P} and \eqref{eq:Eeff} and the relation between $\mathbf{M}$ and $\mathbf{B}$, obtained from Eq.~\eqref{eq:B}, together with Eqs.~\eqref{eq:M} and \eqref{eq:Beff}, we obtain the well-known Clausius-Mossotti relation and its magnetic analog, given by \cite{Panofsky1962,Jackson1999,Bottcher1993,Stratton1941}
\begin{equation}
 \frac{\varepsilon_\mathrm{r}-1}{\varepsilon_\mathrm{r}+2}=\frac{n_\mathrm{a}\alpha_\mathrm{e}}{3\varepsilon_0},
 \hspace{0.5cm}\varepsilon_\mathrm{r}=1+\frac{n_\mathrm{a}\alpha_\mathrm{e}/\varepsilon_0}{1-\frac{n_\mathrm{a}\alpha_\mathrm{e}}{3\varepsilon_0}}
 \label{eq:CMe}
\end{equation}
\begin{equation}
 \frac{\mu_\mathrm{r}-1}{\mu_\mathrm{r}+2}=\frac{\mu_0n_\mathrm{a}\alpha_\mathrm{m}}{3},
 \hspace{0.5cm}\mu_\mathrm{r}=1+\frac{\mu_0n_\mathrm{a}\alpha_\mathrm{m}}{1-\frac{\mu_0n_\mathrm{a}\alpha_\mathrm{m}}{3}}.
 \label{eq:CMm}
\end{equation}
For anisotropic materials, for which Eqs.~\eqref{eq:Eeff} and \eqref{eq:Beff} do not apply, the relations of the permittivity and permeability in Eqs.~\eqref{eq:CMe} and \eqref{eq:CMm} also become more complicated \cite{Aspnes1982}. For dielectric materials with $\varepsilon_\mathrm{r}=n^2$, the Clausius-Mossotti relation in Eq.~\eqref{eq:CMe} is also known as the Lorentz-Lorenz relation \cite{Kragh2018,Bottcher1992}. The Lorentz-Lorenz relation is generally applicable to a wide range of photonic materials as such. However, to fine tune this relation to account for more detailed wavelength, temperature, and density-dependent characteristics of specific materials, experimental parametrizations have been presented so that the right hand side of Eq.~\eqref{eq:CMe} is replaced by a parametrized sum of the wavelength, temperature, and pressure-dependent terms \cite{Schiebener1990,Harvey1998}.

\section{\label{sec:forcedensity}Force density}

Next, we consider different parts of the force density. The force density $\mathbf{f}$ and the field+interaction subsystem quantities $W_\mathrm{emi}$, $\mathbf{G}_\mathrm{emi}$, $\boldsymbol{\mathcal{T}}_\mathrm{emi}$ can be split into electromagnetic field and mechanical pressure parts. For the force density, this splitting reads
\begin{equation}
 \mathbf{f}=\mathbf{f}_\mathrm{em}+\mathbf{f}_\mathrm{mech},
 \label{eq:ftot}
\end{equation}
where $\mathbf{f}_\mathrm{em}$ is the total electromagnetic force density and $\mathbf{f}_\mathrm{mech}$ is the mechanical pressure force density. Here we briefly note that $\mathbf{f}_\mathrm{mech}$ is given for materials with negligible shear strain, e.g., for liquids and gases, by \cite{Kittel2005}
\begin{equation}
 \mathbf{f}_\mathrm{mech}=-\nabla p_\mathrm{mech}.
 \label{eq:fpres}
\end{equation}
Here $p_\mathrm{mech}$ is the mechanical pressure. The mechanical pressure is given in terms of the atomic position field $\mathbf{r}_\mathrm{a}$ of a homogeneous material as $p_\mathrm{mech}=-K\boldsymbol{\nabla}\cdot\mathbf{r}_\mathrm{a}$ \cite{Kittel2005}. The general elastic force density, from which Eq.~\eqref{eq:fpres} is a special case, is obtained by using the elasticity tensor of the specific material as described in Sec.~\ref{sec:elasticitytheory} below.

\subsection{Electromagnetic force density}

The total electromagnetic force density $\mathbf{f}_\mathrm{em}$, appearing in Eq.~\eqref{eq:ftot}, can be split into two parts as
\begin{equation}
 \mathbf{f}_\mathrm{em}=\mathbf{f}_\mathrm{owm}+\mathbf{f}_\mathrm{ost}.
 \label{eq:fem}
\end{equation}
The optical wave momentum force density $\mathbf{f}_\mathrm{owm}$ acts, in the laboratory frame, along the propagation direction of light, and it is responsible for carrying the wave-momentum of light \cite{Bliokh2022,Partanen2022b}. This force density acts on the material atoms at interfaces in such a way that the interface takes the difference of the wave momentum of light in the material and the momentum of light in vacuum. In the bulk, $\mathbf{f}_\mathrm{owm}$ accelerates and decelerates atoms in such a way that the atoms carry part of the total wave momentum of light, and the rest energy of the atomic mass density wave needed for the relativistic covariance of the theory \cite{Partanen2017c,Partanen2019a,Partanen2019b,Partanen2021b}. In contradistinction, the optostrictive force density $\mathbf{f}_\mathrm{ost}$ is an electromagnetic force density acting between the constituents of the material. Therefore, it has a character of action and reaction between the material constituents, and its volume integral including the interface contribution is, thus, at any time equal to zero. Although $\mathbf{f}_\mathrm{ost}$ can give rise to nonzero momentum densities, the total volume integral of the momentum density caused by $\mathbf{f}_\mathrm{ost}$ is zero, and therefore, it does not carry net momentum. In a recent experiment of Astrath \emph{et al.} \cite{Astrath2022}, acoustic waves caused by the radial component of $\mathbf{f}_\mathrm{ost}$ due to a light pulse were discovered.

The accurate position- and time-dependent form of $\mathbf{f}_\mathrm{owm}$ has been described in detail, e.g., in Ref.~\cite{Partanen2022b}. The electrostrictive and magnetostrictive force densities in $\mathbf{f}_\mathrm{ost}$ are both experimentally and theoretically widely known only in the thermodynamical and static field limits. The thermodynamical theory of electrostriction and magnetostriction cannot be applied to optical fields. Therefore, in the present work, we focus in finding the exact position- and time-dependent form of $\mathbf{f}_\mathrm{ost}$ in the optical regime. The total force density in Eq.~\eqref{eq:ftot}, consisting of the terms discussed in the present work, explains the available experimental results for the forces at liquid interfaces \cite{Ashkin1973,Astrath2014,Casner2001}, for the force on a mirror immersed in a liquid \cite{Jones1978,Jones1954}, and for the radial optostrictive force component inside the bulk liquid \cite{Astrath2022}.

\subsubsection{Optical wave momentum force density}

In the rest frame of a nondispersive material, the optical wave momentum force density $\mathbf{f}_\mathrm{owm}$ is given by the Abraham stress-energy-momentum tensor model \cite{Partanen2017c,Milonni2010}. It can be derived from the conservation laws by requiring that the coupled field-material state of light satisfies the covariance principle of the special theory of relativity \cite{Partanen2017c}. In the rest frame of the material, $\mathbf{f}_\mathrm{owm}$ is given by
\begin{equation}
 \mathbf{f}_\mathrm{owm}=-\frac{1}{2}\varepsilon_0|\mathbf{E}|^2\nabla\varepsilon_\mathrm{r}
 -\frac{1}{2}\mu_0|\mathbf{H}|^2\nabla\mu_\mathrm{r}
 +\frac{n^2-1}{c^2}\frac{\partial}{\partial t}(\mathbf{E}\times\mathbf{H}).
 \label{eq:fopt}
\end{equation}
The first two terms of $\mathbf{f}_\mathrm{owm}$ represent the interface force density and the last term is the Abraham volume force density, which has an effect in the direction of the propagation of light. The interface force density has been experimentally verified in several works \cite{Ashkin1973,Astrath2014,Casner2001,Jones1978,Jones1954}, while the Abraham volume force density still lacks direct compelling experimental verification at optical frequencies. The Abraham volume force density, however, has experimental support in the quasistatic limit \cite{Walker1975}. There are works reporting measurements of the Abraham force at optical frequencies \cite{Choi2017,She2008,Kundu2017}, but the interpretation of the results of these measurements is nontrivial \cite{Brevik2009,Mansuripur2009b,Partanen2021a,Brevik2018a,Brevik2018b,Partanen2019e}.

\subsubsection{\label{sec:fst}Optostrictive force density}

Next, we investigate the optostrictive force density $\mathbf{f}_\mathrm{ost}$, which originates from the atomic density dependence of the energy density of the electric and magnetic fields. It can be calculated using the principle of virtual work as discussed in Sec.~\ref{sec:optostriction} below, but here we present first a concise Lorentz-force-law-based derivation of the optostrictive force density. According to Eq.~\eqref{eq:fem}, $\mathbf{f}_\mathrm{ost}=\mathbf{f}_\mathrm{em}-\mathbf{f}_\mathrm{owm}$. We know that the force density $\mathbf{f}_\mathrm{em}$ can be derived from the Lorentz force law as discussed in Appendix \ref{apx:fem}, and the expression of $\mathbf{f}_\mathrm{owm}$ is given in Eq.~\eqref{eq:fopt}. Thus, by using Eqs.~\eqref{eq:fopt} and \eqref{eq:Lorentz2}, we obtain
\begin{equation}
 \mathbf{f}_\mathrm{ost}=\mathbf{f}_\mathrm{oes}+\mathbf{f}_\mathrm{oms}.
 \label{eq:fst}
\end{equation}
where the optoelectro- and optomagnetostrictive force densities $\mathbf{f}_\mathrm{oes}$ and $\mathbf{f}_\mathrm{oms}$ are given by
\begin{equation}
 \mathbf{f}_\mathrm{oes}=\frac{1}{2}\nabla(\mathbf{P}\cdot\mathbf{E}),
 \label{eq:foes}
\end{equation}
\begin{equation}
 \mathbf{f}_\mathrm{oms}=\frac{1}{2}\nabla(\mathbf{M}\cdot\mathbf{B}).
 \label{eq:foms}
\end{equation}

It is conventional to define the optoelectro- and optomagnetostrictive pressures $p_\mathrm{oes}$ and $p_\mathrm{oms}$ related to the force densities $\mathbf{f}_\mathrm{oes}$ and $\mathbf{f}_\mathrm{oms}$ in Eqs.~\eqref{eq:foes} and \eqref{eq:foms} by $\mathbf{f}_\mathrm{oes}=-\nabla p_\mathrm{oes}$ and $\mathbf{f}_\mathrm{oms}=-\nabla p_\mathrm{oms}$. Thus, these pressures can be written as
\begin{equation}
 p_\mathrm{oes}=-\frac{1}{2}\mathbf{P}\cdot\mathbf{E}
 =-\frac{1}{2}\varepsilon_0(\varepsilon_\mathrm{r}-1)|\mathbf{E}|^2,
 \label{eq:poes}
\end{equation}
\begin{equation}
 p_\mathrm{oms}=-\frac{1}{2}\mathbf{M}\cdot\mathbf{B}
 =-\frac{1}{2}\mu_0\mu_\mathrm{r}(\mu_\mathrm{r}-1)|\mathbf{H}|^2.
 \label{eq:poms}
\end{equation}
As discussed in the next section, the compressive work done by the optoelectro- and optomagnetostrictive force densities in Eqs.~\eqref{eq:foes} and \eqref{eq:foms} on the material is found to be less than the change of the field energy. Therefore, we will introduce dissipation terms in the calculation of the optostrictive force density using the principle of virtual work. This is a fundamental change in the conventional way of calculating the electrostrictive and magnetostrictive force densities.

\subsection{\label{sec:optostriction}Virtual work approach to the optostrictive force density}

Next, we investigate the optostrictive force density based on the \emph{principle of virtual work} \cite{Penfield1967,WangN2018}. According to the principle of virtual work, the atoms tend to convert the atomic-density-dependent part of the energy of the field into kinetic and strain energies of the material. The atomic forces following from this principle of virtual work can always be understood as force pairs between atoms, and the resulting total force on the material is a sum of such atomic force pairs. Since the sum of any momentum impulses resulting from pairs of opposite forces is zero, the force density $\mathbf{f}_\mathrm{ost}$ cannot lead to transfer of net momentum in the material, in contrast to the optical wave momentum force density $\mathbf{f}_\mathrm{owm}$ in Eq.~\eqref{eq:fopt}.

The energy density of the electromagnetic field depends on the density of the material through the permittivity and permeability. Depending on which of the fields $\mathbf{D}$, $\mathbf{B}$, $\mathbf{E}$, and $\mathbf{H}$ are kept independent variables during the compression or expansion of the material, the optostrictive force densities $\mathbf{f}_\mathrm{oes}$ and $\mathbf{f}_\mathrm{oms}$ can be calculated from the electric and magnetic energy densities or from their Legendre transforms \cite{Landau1984}
\begin{equation}
 F_\mathrm{e}(\mathbf{D})=W_\mathrm{e},\hspace{0.5cm}
 \tilde F_\mathrm{e}(\mathbf{E})=F_\mathrm{e}-\mathbf{E}\cdot\mathbf{D}=-W_\mathrm{e},
 \label{eq:Fe}
\end{equation}
\begin{equation}
 F_\mathrm{m}(\mathbf{B})=W_\mathrm{m},\hspace{0.5cm}
 \tilde F_\mathrm{m}(\mathbf{H})=F_\mathrm{m}-\mathbf{H}\cdot\mathbf{B}=-W_\mathrm{m},
 \label{eq:Fm}
\end{equation}
in analogy to how the elastic force density is calculated from the strain energy density of the material \cite{Kittel2005}. Here $W_\mathrm{e}=\frac{1}{2}\varepsilon_0\varepsilon_\mathrm{r}|\mathbf{E}|^2=\frac{1}{2\varepsilon_0\varepsilon_\mathrm{r}}|\mathbf{D}|^2$ is the energy density of the electric field and $W_\mathrm{m}=\frac{1}{2}\mu_0\mu_\mathrm{r}|\mathbf{H}|^2=\frac{1}{2\mu_0\mu_\mathrm{r}}|\mathbf{B}|^2$ is the energy density of the magnetic field. The independent variables in Eqs.~\eqref{eq:Fe} and \eqref{eq:Fm} are indicated by parentheses. Conventionally, in the thermodynamical derivation of the electro- and magnetostrictive force densities \cite{Landau1984}, Eqs.~\eqref{eq:Fe} and \eqref{eq:Fm} correspond to the field-dependent parts of the free energy densities of the system. In previous literature, the thermodynamical derivation is considered to be extended to time-dependent fields as such or by considering the time-averages of free energy densities over the harmonic cycle \cite{Landau1984,Boyd2008,Brevik2018a}. For a discussion of the electro- and magnetostriction for stationary fields, see Appendix \ref{apx:stationary}.

In the case of an optical field, the compression or expansion of the material associated with optostriction takes place simultaneously with the flow of energy and momentum into direction of the propagation of light. Therefore, we can define boundary conditions related to energy and momentum fluxes in the calculation of the optostrictive force density, which are different from the boundary conditions used in Eqs.~\eqref{eq:Fe} and \eqref{eq:Fm}, where we keep certain fields constant. We do not elaborate these flux-based boundary conditions further in this work.

In the following, we propose that the changes of the dipole moments $\mathbf{p}$ and $\mathbf{m}$ by the atomic number density variation $\delta n_\mathrm{a}$, in the time-dependent case, introduce a hitherto unknown dissipation mechanism to be accounted for. This dissipation can, for example, correspond to radiation loss and be related to non-conservativity of time-dependent optical forces \cite{Sukhov2017}. We determine its contribution by requiring that the sum of the work done by the Lorentz-force-law-based optostrictive force density $\mathbf{f}_\mathrm{ost}$ in Eq.~\eqref{eq:fst} and the dissipation terms is equal to the change of the field energy density. Consequently, in the calculation of the force densities below, from the full differentials $(\delta\tilde F_\mathrm{e})_\mathbf{E}=-(\delta W_\mathrm{e})_\mathbf{E}=-\frac{1}{2}\varepsilon_0\frac{\partial\varepsilon_\mathrm{r}}{\partial n_\mathrm{a}}|\mathbf{E}|^2\delta n_\mathrm{a}$ and $(\delta F_\mathrm{m})_\mathbf{B}=(\delta W_\mathrm{m})_\mathbf{B}=-\frac{1}{2\mu_0\mu_\mathrm{r}^2}\frac{\partial\mu_\mathrm{r}}{\partial n_\mathrm{a}}|\mathbf{B}|^2\delta n_\mathrm{a}$, for fixed $\mathbf{E}$ and $\mathbf{B}$, we subtract the dissipation-related differentials  $(\delta W_\mathrm{e,diss})_\mathbf{E}=-\frac{1}{2}n_\mathrm{a}\big(\frac{\partial\mathbf{p}}{\partial n_\mathrm{a}}\big)_\mathbf{E}\cdot\mathbf{E}\delta n_\mathrm{a}$ and $(\delta W_\mathrm{m,diss})_\mathbf{B}=-\frac{1}{2}n_\mathrm{a}\big(\frac{\partial\mathbf{m}}{\partial n_\mathrm{a}}\big)_\mathbf{B}\cdot\mathbf{B}\delta n_\mathrm{a}$. Thus, we obtain the dissipation-reduced differentials as $(\delta\tilde F_\mathrm{e}')_\mathbf{E}=(\delta\tilde F_\mathrm{e})_\mathbf{E}-(\delta W_\mathrm{e,diss})_\mathbf{E}$ and $(\delta F_\mathrm{m}')_\mathbf{B}=(\delta F_\mathrm{m})_\mathbf{B}-(\delta W_\mathrm{m,diss})_\mathbf{B}$. Therefore, the optoelectro- and optomagnetostrictive force densities are given by
\begin{align}
 \mathbf{f}_\mathrm{oes} &=\sum_{i=1}^3\sum_{j=1}^3\partial_j\Big(\frac{\delta\tilde F_\mathrm{e}'}{\delta(\epsilon_\mathrm{a})_{ij}}\Big)_\mathbf{E}\hat{\mathbf{e}}_{i}\nonumber\\
 & = -\nabla\Big[n_\mathrm{a}\Big(\frac{\delta\tilde F_\mathrm{e}'}{\delta n_\mathrm{a}}\Big)_\mathbf{E}\Big]\nonumber\\
 & =\frac{1}{2}\nabla(n_\mathrm{a}\mathbf{p}\cdot\mathbf{E})\nonumber\\
 & =\frac{1}{2}\nabla\Big(\mathbf{P}\cdot\mathbf{E}\Big),
 \label{eq:foes2}
\end{align}
\begin{align}
 \mathbf{f}_\mathrm{oms} &=\sum_{i=1}^3\sum_{j=1}^3\partial_j\Big(\frac{\delta F_\mathrm{m}'}{\delta(\epsilon_\mathrm{a})_{ij}}\Big)_\mathbf{B}\hat{\mathbf{e}}_{i}\nonumber\\
 & = -\nabla\Big[n_\mathrm{a}\Big(\frac{\delta F_\mathrm{m}'}{\delta n_\mathrm{a}}\Big)_\mathbf{B}\Big]\nonumber\\
 & =\frac{1}{2}\nabla(n_\mathrm{a}\mathbf{m}\cdot\mathbf{B})\nonumber\\
 & =\frac{1}{2}\nabla\Big(\mathbf{M}\cdot\mathbf{B}\Big).
 \label{eq:foms2}
\end{align}
Here $\hat{\mathbf{e}}_i$, $i\in\{x,y,z\}$ are the three unit vectors of the Cartesian coordinate system. The subscripts $\mathbf{E}$ and $\mathbf{B}$ indicate that these fields are taken as constants in the calculation of the differentials. In the second equalities of Eqs.~\eqref{eq:foes2} and \eqref{eq:foms2}, we have assumed that the material is isotropic. In this case, we can convert the derivatives with respect to the components of the atomic position strain tensor $\boldsymbol{\epsilon}_\mathrm{a}=\textstyle\frac{1}{2}[\boldsymbol{\nabla}\otimes\mathbf{r}_\mathrm{a}+(\boldsymbol{\nabla}\otimes\mathbf{r}_\mathrm{a})^T]$, where the superscript $T$ denotes the transpose,  into derivatives with respect to the atomic number density as $\sum_{i=1}^3\sum_{j=1}^3\partial_j\big(\frac{\partial F}{\partial(\epsilon_\mathrm{a})_{ij}}\big)\hat{\mathbf{e}}_{i}=-\nabla(n_\mathrm{a}\frac{\partial F}{\partial n_\mathrm{a}})$ \cite{Sun2015,Landau1984}.

If the derivation of the force densities is done in analogy to Eqs.~\eqref{eq:foes2} and \eqref{eq:foms2} except that the we use the energy functions without the subtraction of the dissipation terms, the optoelectro- and optomagnetostrictive pressures become $p_\mathrm{es}=-\frac{1}{2}\mathbf{P}\cdot\mathbf{E}_\mathrm{eff}$ and $p_\mathrm{ms}=-\frac{1}{2}\mathbf{M}\cdot\mathbf{B}_\mathrm{eff}$. This means that the time-dependent macroscopic fields are replaced by the corresponding local fields. By their definition, these optoelectro- and optomagnetostrictive pressures are inherently energy-conserving during the compression. Using Eqs.~\eqref{eq:Eeff}--\eqref{eq:CMm}, we can relate these pressures to the pressures in Eqs.~\eqref{eq:poes} and \eqref{eq:poms} as $p_\mathrm{es}=\frac{\varepsilon_\mathrm{r}+2}{3}p_\mathrm{oes}$ and $p_\mathrm{ms}=\frac{\mu_\mathrm{r}+2}{3\mu_\mathrm{r}}p_\mathrm{oms}$. The relation of these pressures to experiments is discussed in Sec.~\ref{sec:discussion}.

\subsection{\label{sec:elasticitytheory}Optoelastic strain force density}

Next, we present how the mechanical pressure and optostrictive force densities, $\mathbf{f}_\mathrm{mech}$ and $\mathbf{f}_\mathrm{ost}$, can be obtained simultaneously using the elasticity theory formalism of the total strain. As described below, the force densities of Eqs.~\eqref{eq:fpres} and \eqref{eq:fst} are obtained as special cases of the general force density expressions of this section. The total strain energy density is given by \cite{Kittel2005}
\begin{equation}
 W_\mathrm{strain} =\frac{1}{2}\sum_{i=1}^3\sum_{j=1}^3\sigma_{ij}\epsilon_{ij}.
\end{equation}
Here $\sigma_{ij}$ are elements of the elastic stress tensor $\boldsymbol{\sigma}$ and $\epsilon_{ij}$ are elements of the total strain tensor $\boldsymbol{\epsilon}$. The total strain tensor $\boldsymbol{\epsilon}$ is given by \cite{Kittel2005}
\begin{equation}
 \boldsymbol{\epsilon}=\boldsymbol{\epsilon}_\mathrm{a}-\boldsymbol{\epsilon}_\mathrm{e}-\boldsymbol{\epsilon}_\mathrm{m}.
\end{equation}
Here the atomic strain tensor $\boldsymbol{\epsilon}_\mathrm{a}$ is given by
\begin{equation}
 \boldsymbol{\epsilon}_\mathrm{a}=\textstyle\frac{1}{2}[\boldsymbol{\nabla}\otimes\mathbf{r}_\mathrm{a}+(\boldsymbol{\nabla}\otimes\mathbf{r}_\mathrm{a})^T].
\end{equation}
For the electric and magnetic strain tensors, we adopt the most general description from previous literature \cite{Newnham1997}.
The electric and magnetic strain tensors are given by
\begin{align}
 \boldsymbol{\epsilon}_\mathrm{e} &=\mathbf{Q}_\mathrm{e}^{(1)}:\mathbf{P}+\mathbf{Q}_\mathrm{e}^{(2)}:(\mathbf{P}\otimes\mathbf{P})+\cdots,\nonumber\\
 \boldsymbol{\epsilon}_\mathrm{m} &=\mathbf{Q}_\mathrm{m}^{(1)}:\mathbf{M}+\mathbf{Q}_\mathrm{m}^{(2)}:(\mathbf{M}\otimes\mathbf{M})+\cdots.
 \label{eq:Qem}
\end{align}
The first terms are associated with the piezoelectricity and piezomagnetism and the second terms are associated with the electro- and magnetostriction. Also, higher-order terms can be accounted for. In Eq.~\eqref{eq:Qem}, the quantities $\mathbf{Q}_\mathrm{e}^{(1)}$ and $\mathbf{Q}_\mathrm{m}^{(1)}$ are the piezoelectric and piezomagnetic tensors and $\mathbf{Q}_\mathrm{e}^{(2)}$ and $\mathbf{Q}_\mathrm{m}^{(2)}$ are the electro- and magnetostriction tensors.

Using the elasticity tensor $\mathbf{C}$ of an isotropic material with bulk modulus $K$ and shear modulus $G$, the elastic stress tensor $\boldsymbol{\sigma}$ is given by
\begin{equation}
 \boldsymbol{\sigma}=\mathbf{C}:\boldsymbol{\epsilon}=\textstyle(K-\frac{2}{3}G)\mathrm{Tr}(\boldsymbol{\epsilon})\mathbf{I}+2G\boldsymbol{\epsilon}.
\end{equation}
Here $\mathrm{Tr}(x)$ denotes the trace of a matrix and $\mathbf{I}$ is a $3\times3$ unit matrix. In this case, the total optoelastic strain force density is given by
\begin{align}
 \mathbf{f}_\mathrm{strain} &=\boldsymbol{\nabla}\cdot\boldsymbol{\sigma}
 =\sum_{i=1}^3\sum_{j=1}^3\partial_j\Big(\frac{\partial W_\mathrm{strain}}{\partial(\epsilon_\mathrm{a})_{ij}}\Big)\hat{\mathbf{e}}_{i}\nonumber\\
  &=\textstyle(K+\frac{4}{3}G)\nabla[\boldsymbol{\nabla}\cdot\mathbf{r}_\mathrm{a}]-G\boldsymbol{\nabla}\times[\boldsymbol{\nabla}\times\mathbf{r}_\mathrm{a}]\nonumber\\
  &\hspace{0.5cm}-\textstyle(K+\frac{4}{3}G)\nabla\mathrm{Tr}(\boldsymbol{\epsilon}_\mathrm{e}+\boldsymbol{\epsilon}_\mathrm{m})-2G\boldsymbol{\nabla}\cdot(\boldsymbol{\epsilon}_\mathrm{e}+\boldsymbol{\epsilon}_\mathrm{m}).
  \label{eq:fstrain}
\end{align}

In our special case of a material with no shear strain, i.e., a liquid or gas, we can set $G=0$. We assume no piezoelectricity and -magnetism, use the conventional constitutive relations of the fields in a nondispersive material, and give for the fourth-rank optical electro- and magnetostriction tensors the values
\begin{equation}
 \mathbf{Q}_\mathrm{e}^{(2)}=-\frac{\mathbf{I}}{2K\varepsilon_0(\varepsilon_\mathrm{r}-1)},
 \hspace{0.5cm}
 \mathbf{Q}_\mathrm{m}^{(2)}=-\frac{\mu_0\mu_\mathrm{r}\mathbf{I}}{2K(\mu_\mathrm{r}-1)},
\end{equation}
where $\mathbf{I}$ is the identity operator for the matrices $\mathbf{P}\otimes\mathbf{P}$ and $\mathbf{M}\otimes\mathbf{M}$ in Eq.~\eqref{eq:Qem}. Thus, we obtain $\boldsymbol{\epsilon}_\mathrm{e}=-\frac{1}{2K}\mathbf{P}\otimes\mathbf{E}$, $\boldsymbol{\epsilon}_\mathrm{m}=-\frac{1}{2K}\mathbf{M}\otimes\mathbf{B}$, and consequently Eq.~\eqref{eq:fstrain} becomes
\begin{equation}
 \mathbf{f}_\mathrm{strain}=\mathbf{f}_\mathrm{mech}+\mathbf{f}_\mathrm{ost},
 \label{eq:fstrain2}
\end{equation}
where $\mathbf{f}_\mathrm{mech}$ is equal to the expression given in Eq.~\eqref{eq:fpres} and $\mathbf{f}_\mathrm{ost}$ is equal to the expression given in Eq.~\eqref{eq:fst}, with the optoelectro- and optomagnetostrictive force densities given in Eqs.~\eqref{eq:foes} and \eqref{eq:foms}.

\begin{figure*}
\centering
\includegraphics[width=\textwidth]{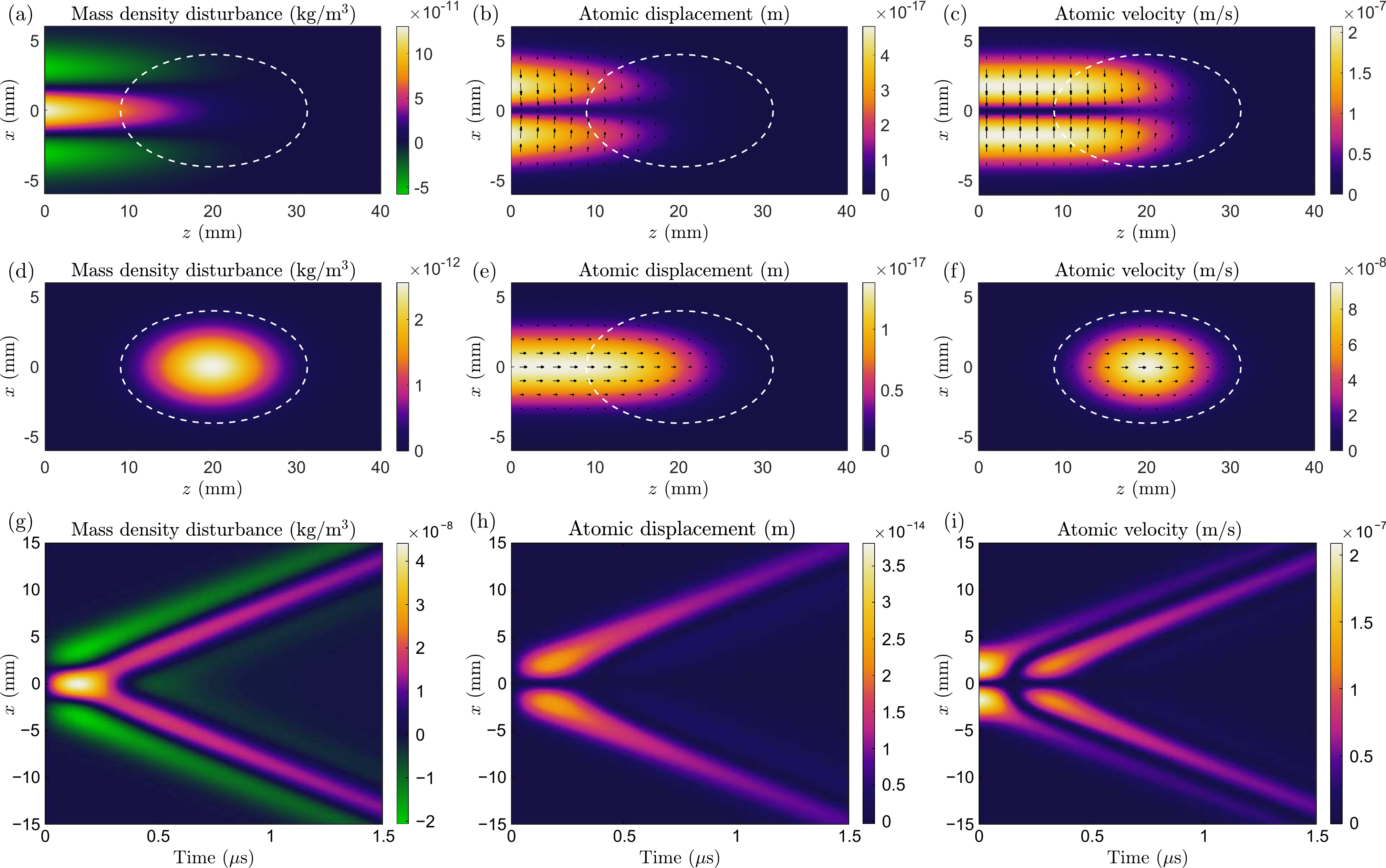}
\caption{\label{fig:simupulse}
Simulation of a Gaussian light pulse propagating along the positive $z$ axis marked by a dashed line ellipse centered at position $z=20$ mm (a--f). The panels represent (a) the mass density disturbance, (b) atomic displacement, and (c) atomic velocity resulting from the total force density $\mathbf{f}=\mathbf{f}_\mathrm{owm}+\mathbf{f}_\mathrm{ost}$ of a Gaussian light pulse in silicon. The \emph{optoelectrostrictive} force density $\mathbf{f}_\mathrm{ost}$ is the \emph{dominating} force density in these figures. Panels (d)--(f) show the mass density disturbance, atomic displacement, and atomic velocity contributions resulting from $\mathbf{f}_\mathrm{owm}$. Simulation of optically generated elastic waves after the light pulse has passed is presented in panels (g)--(i). These panels show the time dependence of (g) the mass density disturbance, (h) atomic displacement, and (i) atomic velocity at $z=0$ mm. The colorbars show the magnitudes, and the arrows show the directions of the vector quantities.}
\end{figure*}

\section{\label{sec:simulations}Simulation of optical electrostriction in silicon}

Next, we study the implications of the theory above by simulating the electrostrictive effect using first a Gaussian light pulse and second a continuous-wave Gaussian light beam. We have selected silicon as an example of an electrostrictive photonic material for which we carry out the numerical simulations. The refractive index of silicon is $n=3.4757$ for the selected vacuum wavelength of $\lambda_0=1550$ nm \cite{Li1980}. The density of silicon is $\rho_\mathrm{a}=2329$ kg/m$^3$ \cite{Lide2004}, the bulk modulus is $K=97.8$ GPa, and the shear modulus is $G=79.6$ GPa \cite{Hopcroft2010}. These values correspond to the compressibility of $C=1/(K+\frac{4}{3}G)=4.90\times10^{-12}$ Pa$^{-1}$. In the present work, we do not account for the elastic anisotropy of the silicon crystal, but model it using the scalar elastic parameters above, in the case of which the elastic strain force density is given by the first two terms of Eq.~\eqref{eq:fstrain} \cite{Mavko2003,Partanen2017e,Landau1975,Kittel2005}.

In more accurate simulations, the elastic parameters $K$ and $G$ and the associated elastic force density in Eq.~\eqref{eq:fstrain} can be replaced by the use of the complete elasticity matrix and the corresponding elastic force density \cite{Kittel2005}. For silicon, and more generally for solids and liquids, which are hard to compress, the elastic parameters and derivatives with respect to the density are very accurately equal for isothermal and isentropic processes. Thus, we can neglect thermal effects in the studies of electrostrictive compression in low-loss materials, such as silicon at the wavelength used in our simulations.

\subsection{Gaussian light pulse}
Next, we perform simulations for a continuous-wave Gaussian light beam. The electric field of a one-dimensional Gaussian light pulse \cite{Griffiths1998}, linearly polarized in the direction of the $x$ axis and propagating along the positive $z$ axis in silicon, is given in cylindrical coordinates $\mathbf{r}=(r,\phi,z)$ by
\begin{equation}
 \mathbf{E}(\mathbf{r},t)=E_0e^{-r^2/w_0^2}\cos[kz-\omega t]e^{-(\Delta k)^2(z-ct/n)^2/2}\hat{\mathbf{e}}_x.
\end{equation}
Here $E_0$ is the electric field amplitude, $w_0$ is the beam waist radius at which the intensity drops to $1/e^2$ of its axial value, $k=n\omega/c=2\pi n/\lambda_0$ is the wave number, and $\Delta k$ is the standard deviation of the wave number, for which we use $\Delta k=10^{-5}k$. For the beam waist radius $w_0$, we use the value of $w_0=3.489$ mm, and for the electric field amplitude $E_0$, we use the value of $E_0=1.972\times10^7$ V/m. These values correspond to the total electromagnetic energy of the pulse equal to $U_0=\varepsilon_0\varepsilon_\mathrm{r}E_0^2\pi^{3/2}w_0^2/(4\Delta k)=5.00$ mJ. The magnetic field corresponding to the electric field above is determined by Maxwell's equations.

Figure \ref{fig:simupulse}(a) shows the atomic mass density disturbance
resulting from the total force density of a Gaussian light pulse in silicon at the instant of time when the center of the Gaussian light pulse propagating along the positive $z$ axis is at $z=20$ mm. The mass density disturbance is spatially averaged over the harmonic cycle. It is seen that the mass density disturbance has positive values at the $z$ axis behind the light pulse ($z<20$ mm, $|y|<2$ mm), and negative values are obtained away from the $z$ axis ($z<20$ mm, 2 mm $<|y|<6$ mm). The corresponding atomic displacement distribution is depicted in Fig.~\ref{fig:simupulse}(b) and the atomic velocity distribution is shown in Fig.~\ref{fig:simupulse}(c). The radial component of the optostrictive force density $\mathbf{f}_\mathrm{ost}$ dominates over the optical wave momentum force density $\mathbf{f}_\mathrm{owm}$ and the longitudinal component of $\mathbf{f}_\mathrm{ost}$ in producing the atomic distributions in Figs.~\ref{fig:simupulse}(a)--\ref{fig:simupulse}(c). The atomic mass density disturbance, atomic displacement, and atomic velocity components following from $\mathbf{f}_\mathrm{owm}$ are depicted in Figs.~\ref{fig:simupulse}(d)--\ref{fig:simupulse}(f). These distributions are fractions of the total distributions in Figs.~\ref{fig:simupulse}(a)--\ref{fig:simupulse}(c).

Figures \ref{fig:simupulse}(g)--\ref{fig:simupulse}(i) show the time-dependence of the elastic relaxation of the mass density disturbance, atomic displacement, and atomic velocity by elastic waves at $z=0$ mm after the light pulse has passed. The distributions at $t=0$ $\mu$s correspond to the quantities of Figs.~\ref{fig:simupulse}(a)--\ref{fig:simupulse}(c) behind the light pulse at $z=0$ mm. The relaxation takes place at the velocity of sound after the light pulse has passed. The sound waves, caused by the optoelectrostrictive force density, propagate radially outward from the beam axis ($x=0$ mm). The displacement of the material by the optostrictive force density takes place also parallel to the beam axis. The relaxation of this displacement is not depicted in Fig.~\ref{fig:simupulse}. The emergence of the mass density disturbance and the related elastic waves have been experimentally detected for a light pulse in water \cite{Astrath2022}. The detection of optostrictively produced elastic waves in the present example case of silicon should also be experimentally feasible.

\begin{figure}
\centering
\includegraphics[width=\columnwidth]{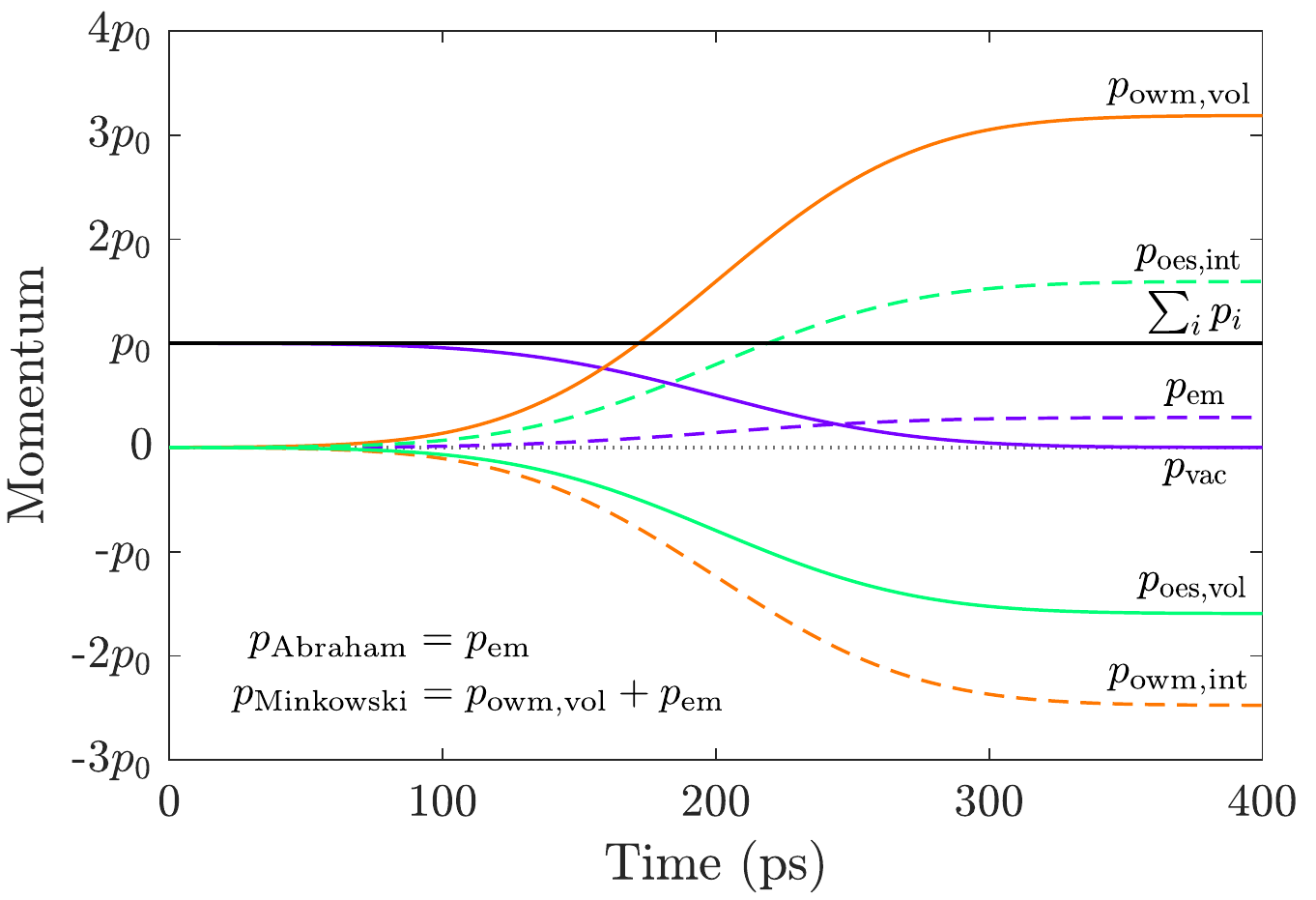}
\caption{\label{fig:curves}
Unified picture of the momentum components for a normally incident Gaussian light pulse crossing a vacuum-silicon interface with an antireflective coating. The center of the pulse is at the position of the interface at $t=200$ ps. The normal components of the momenta are given in units of $p_0=U_0/c$, the total momentum of the pulse in vacuum. The component $p_\mathrm{vac}$ is the momentum of the part of the pulse located in vacuum, and $p_\mathrm{em}$ is the momentum of the electromagnetic field in silicon.  The material momentum driven by $\mathbf{f}_\mathrm{owm}$ in the bulk silicon is denoted by $p_\mathrm{owm,vol}$, and the material momentum generated by $\mathbf{f}_\mathrm{owm}$ at the interface is denoted by $p_\mathrm{owm,int}$. Correspondingly, the material momentum driven by $\mathbf{f}_\mathrm{oes}$ in the bulk silicon is denoted by $p_\mathrm{oes,vol}$, and the material momentum generated by $\mathbf{f}_\mathrm{owm}$ at the interface is denoted by $p_\mathrm{oes,int}$. Note that the radial momentum densities integrate out in this figure, which represents only the total volume integrated values.}
\end{figure}

Figure \ref{fig:curves} presents the momentum components that follow from our unified optical force theory when a Gaussian light pulse crosses a vacuum-silicon interface with an antireflective coating. For simplicity, the pulse is assumed to propagate along the surface normal and the antireflective coating is assumed to be perfect. The center of the pulse reaches the interface at $t=200$ ps. At $t=0$ ps, the pulse is entirely in vacuum, its momentum is $p_\mathrm{vac}=p_0=U_0/c$, and the other momentum components are zero.  When the pulse has fully crossed the interface, the momentum of the electromagnetic field part of the pulse is $p_\mathrm{em}=p_0/n$, which is equal to the \emph{Abraham momentum} of light. The atomic mass density wave, driven by the optical wave momentum force density $\mathbf{f}_\mathrm{owm}$, has momentum $p_\mathrm{owm,vol}=(n-1/n)p_0$, which together with the electromagnetic momentum component $p_\mathrm{em}=p_0/n$ gives the total wave momentum of light. The total wave momentum $p_\mathrm{owm,vol}+p_\mathrm{em}=np_0$ is equal to the \emph{Minkowski momentum} of light. In the bulk silicon, the optoelectrostrictive force density $\mathbf{f}_\mathrm{oes}$ gives rise to the momentum component equal to $p_\mathrm{oes,vol}=-\frac{1}{2}(n-1/n)p_0$. The material interface takes the momentum component $p_\mathrm{owm,int}=(1-n)p_0$ from $\mathbf{f}_\mathrm{owm}$ and $p_\mathrm{oes,int}=\frac{1}{2}(n-1/n)p_0$ from $\mathbf{f}_\mathrm{oes}$. The sum of all momentum components is equal to $p_0$ \emph{at all times}. Since moreover $p_\mathrm{oes,vol}+p_\mathrm{oes,int}=0$ at all times, the optoelectrostriction does not lead to transfer of volume integrated net momentum.

In the mass-polariton theory of light \cite{Partanen2017c,Partanen2019a}, the momentum of the atomic mass density wave is given, using the notation of the present work, by $p_\mathrm{owm,vol}$ and the momentum of the mass-polariton state of light is correspondingly $p_\mathrm{MP}=p_\mathrm{owm,vol}+p_\mathrm{em}=np_0$. This is the refractive-index-proportional momentum of light in a material, which is observed in most experiments \cite{Jones1954,Jones1978,Campbell2005}. The present theory is, thus, in full agreement with the covariant quasiparticle properties of the mass-polariton theory \cite{Partanen2017c,Partanen2019a,Partanen2021b}.

\begin{figure*}
\centering
\includegraphics[width=\textwidth]{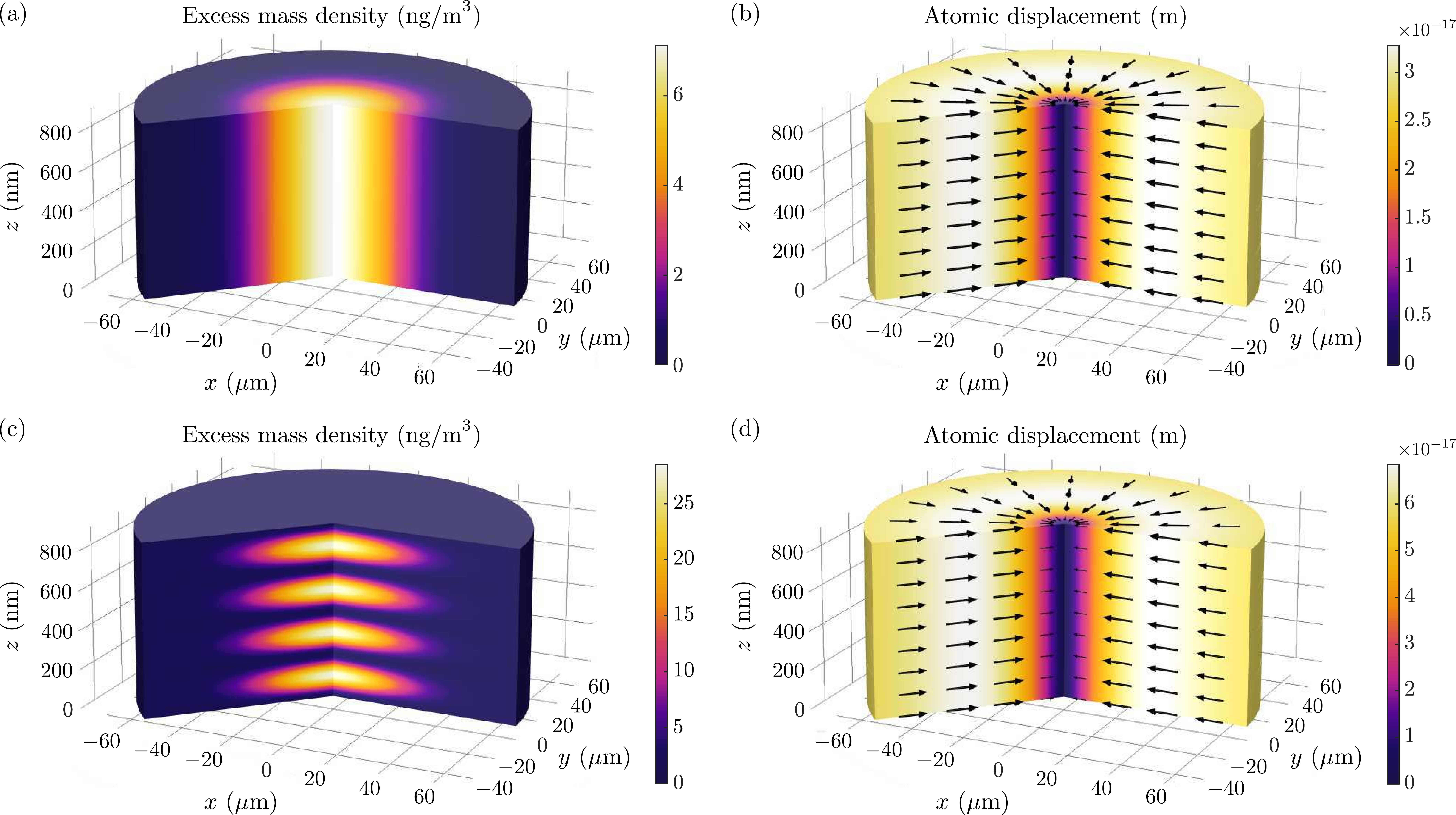}
\caption{\label{fig:simulation}
Simulations of optoelectrostriction for a radially Gaussian continuous-wave light beam. (a) The excess mass density and (b) the atomic displacements for a light beam propagating along the $z$ axis in a homogeneous silicon crystal. (c) The excess mass density and (d) the atomic displacements for the standing wave of a light beam incident from silicon along the negative $z$ axis and reflected from a perfect mirror at $z=0$ nm. The figures show the quantities for the length of two wavelengths in the cylindrical central region of the simulation geometry. The colors show the magnitude and the arrows show the direction of the vector quantities.}
\end{figure*}

\subsection{Continuous-wave Gaussian light beam}
The electric field of an incident continuous-wave Gaussian light beam propagating along the negative $z$ axis, polarized along the $x$ axis, and focused to $x=y=z=0$ is given in cylindrical coordinates by \cite{Novotny2006}
\begin{align}
 &\mathbf{E}(\mathbf{r},t)\nonumber\\
 &=E_0\frac{w_0}{w(z)}e^{-r^2/w^2(z)}\cos\!\Big[kz+\omega t+\frac{kr^2}{2R(z)}-\varphi(z)\Big]\hat{\mathbf{e}}_x.
 \label{eq:Efield}
\end{align}
Here $w(z)=w_0\sqrt{1+(z/z_\mathrm{R})^2}$ is the position-dependent beam radius, $z_\mathrm{R}=\pi w_0^2n/\lambda_0$ is the Rayleigh range, $R(z)=z[1+(z_\mathrm{R}/z)^2]$ is the radius of curvature of the wavefronts, and $\varphi(z)=\arctan(z/z_\mathrm{R})$ is the Gouy phase. The electric field amplitude at the focus is given by
$E_0=\sqrt{4P_0n/(\pi w_0^2c\varepsilon_0\varepsilon_\mathrm{r})}$, where $P_0$ is the average power of the beam. In the simulation of the continuous-wave beam, for the beam waist radius $w_0$, we use the value of $w_0=40$ $\mu$m, and for the average power of the beam, we use $P_0=1$ W.

In order to obtain a stationary atomic distribution as a result of the simulation, in Newton's equation we add the force density term $\mathbf{f}_\mathrm{damping}=-\Gamma\rho_\mathrm{a}\mathbf{v}_\mathrm{a}$, which is associated with the damping of mechanical waves \cite{MaartenvanDoorn2020}, quantified here by the damping frequency $\Gamma$. In the case of a continuous-wave field, the damping frequency determines the time scale at which the excess mass density and atomic displacement distributions studied below are formed, but it does not affect the values of the final distributions.

Figure \ref{fig:simulation}(a) presents the excess mass density resulting from the electrostriction induced by the Gaussian light beam of Eq.~\eqref{eq:Efield} in a homogeneous silicon crystal. The field is only slightly focused, and thus, the variations in the mass density along the $z$ axis are not visible, while in the transverse plane, the excess mass density follows the Gaussian form of the light beam. Apart from very small variations, the result is independent of whether the square of the electric field used in the simulations is averaged over the harmonic cycle or not. The atomic displacement field corresponding to the excess mass density of Fig.~\ref{fig:simulation}(a) is presented in Fig.~\ref{fig:simulation}(b). This displacement field is radially directed toward the beam axis, where the displacement field is zero. Far from the beam axis, the atomic displacement field asymptotically approaches zero again. This takes place outside the region depicted in the figure.

Figure \ref{fig:simulation}(c) shows the excess mass density resulting from the electrostriction induced by the Gaussian light beam of Eq.~\eqref{eq:Efield} when the beam is reflected from a perfect mirror positioned at $z=0$ nm. The excess mass density follows the standing wave pattern of the square of the electric field as expected. Due to the reflection and the standing wave pattern, the highest values of the excess mass density in Fig.~\ref{fig:simulation}(c) are four times the highest values the same quantity for the Gaussian light beam without reflection from the mirror in Fig.~\ref{fig:simulation}(a). The atomic displacement field corresponding to the excess mass density of Fig.~\ref{fig:simulation}(c) is depicted in Fig.~\ref{fig:simulation}(d). Due to the standing wave pattern, the atomic displacement field associated with electrostriction has an additional nonradial contribution. This component is, however, so small that it is not visible in Fig.~\ref{fig:simulation}(d).

\section{\label{sec:tensors}Stress-energy-momentum tensors}

The SEM tensor of a physical system or a subsystem compiles the energy and momentum densities and the stress tensor of the conservation laws, exemplified by Eqs.~\eqref{eq:momentumconservation}--\eqref{eq:conservationphi}, in a single second-rank physical quantity. The contravariant form of an arbitrary SEM tensor in the Minkowski space-time is defined by
$\mathbf{T}=T^{\alpha\beta}\mathbf{e}_\alpha\otimes\mathbf{e}_\beta$, where the Einstein summation convention is used, and $\mathbf{e}_\alpha$ and $\mathbf{e}_\beta$ are unit vectors of the four-dimensional space-time, i.e., $(ct,x,y,z)$. The Greek indices range over the dimensions of the space-time. The matrix representation of $\mathbf{T}$ is given by \cite{Landau1989,Jackson1999,Misner1973,Partanen2022b}
\begin{equation}
 \mathbf{T}=
 \left[\begin{array}{cc}
  W & c\mathbf{G}^T\\
  c\mathbf{G} & \boldsymbol{\mathcal{T}}\\
 \end{array}\right]
 =\left[\begin{array}{cccc}
  W & cG^x & cG^y & cG^z\\
  cG^x & \mathcal{T}^{xx} & \mathcal{T}^{xy} & \mathcal{T}^{xz}\\
  cG^y & \mathcal{T}^{yx} & \mathcal{T}^{yy} & \mathcal{T}^{yz}\\
  cG^z & \mathcal{T}^{zx} & \mathcal{T}^{zy} & \mathcal{T}^{zz}
 \end{array}\right].
 \label{eq:emt}
\end{equation}
In some previous literature, asymmetric SEM tensors have been introduced \cite{Penfield1967}, but in our work, all SEM tensors are symmetric and strictly based on the classical definition in Eq.~\eqref{eq:emt}.

We require that the SEM tensors of the total system and its subsystems, e.g., the material and field+interaction subsystems, must all transform between inertial frames in a relativistically covariant way. This imposes strong limitations on the possible forms of the SEM tensors. Since the SEM tensor is a derived quantity made of basic physical quantities, such as the atomic density, atomic velocity, and the electric and magnetic fields, it can be transformed between inertial frames by transforming the basic physical quantities in its elements. Simultaneously, the SEM tensor must satisfy the Lorentz transformation of second-rank tensors \cite{Penfield1967,Kemp2017,Partanen2019a,Partanen2021b}. Thus, arbitrary combinations of basic physical quantities forming the SEM tensor are not possible. In addition, one cannot arbitrarily mix tensor elements between the SEM tensors of two subsystems. Below, the SEM tensors are presented so that they each satisfy the condition discussed above. They are presented in terms of four-dimensional quantities, such as four-vectors, field tensors, and Lorentz scalars, which makes the fulfillment of the relativistic covariance property transparent.

The total SEM tensor of the system of the electromagnetic field and a perfect fluid is given by a sum of the SEM tensor $\mathbf{T}_\mathrm{mat}$ of the material subsystem, and the SEM tensor $\mathbf{T}_\mathrm{emi}$ of the field+interaction subsystem as
\begin{equation}
\mathbf{T}_\mathrm{tot}=\mathbf{T}_\mathrm{mat}+\mathbf{T}_\mathrm{emi}.
\end{equation}
By defining the four-fource density as $\boldsymbol{\mathcal{F}}=(\phi/c,\mathbf{f})$, the conservation laws in Eqs.~\eqref{eq:momentumconservation}--\eqref{eq:conservationphi} are compactly written as $\partial_\beta(T_\mathrm{mat})^{\alpha\beta}=\mathcal{F}^\alpha$ and $\partial_\beta(T_\mathrm{emi})^{\alpha\beta}=-\mathcal{F}^\alpha$. For the total isolated system, the conservation laws are correspondingly given by $\partial_\beta(T_\mathrm{tot})^{\alpha\beta}=0$.

In accordance with the splitting of the force density into parts in previous sections, the SEM tensor of the field+interaction subsystem can be written as
\begin{equation}
\mathbf{T}_\mathrm{emi}=\mathbf{T}_\mathrm{owm}+\mathbf{T}_\mathrm{ost}+\mathbf{T}_\mathrm{mech}.
\end{equation}
Here $\mathbf{T}_\mathrm{owm}$ is the optical generalized Abraham SEM tensor of the field corresponding to $\mathbf{f}_\mathrm{owm}$, $\mathbf{T}_\mathrm{ost}$ is the electro- and magnetostriction SEM tensor corresponding to $\mathbf{f}_\mathrm{ost}$, and $\mathbf{T}_\mathrm{mech}$ is the SEM tensor of mechanical pressure corresponding to $\mathbf{f}_\mathrm{mech}$.

Using the diagonal Minkowski metric tensor $\boldsymbol{g}$ with sign convention $g_{00}=1$, $g_{11}=g_{22}=g_{33}=-1$, the subsystem SEM tensors above are given in contravariant forms applicable to a general inertial frame as
\begin{equation}
 \mathbf{T}_\mathrm{mat}=\rho_\mathrm{a}\mathbf{U}_\mathrm{a}\otimes\mathbf{U}_\mathrm{a},
 \label{eq:Tmat}
\end{equation}
\begin{align}
 \mathbf{T}_\mathrm{owm} &=\frac{1}{2}(\boldsymbol{\mathcal{F}}\boldsymbol{g}\boldsymbol{\mathcal{D}}+\boldsymbol{\mathcal{D}}\boldsymbol{g}\boldsymbol{\mathcal{F}})-\frac{1}{4}\boldsymbol{g}\mathrm{Tr}[\boldsymbol{\mathcal{F}}\boldsymbol{g}\boldsymbol{\mathcal{D}}\boldsymbol{g}]\nonumber\\
 &\hspace{0.4cm}-\frac{1}{2c^2}[(\boldsymbol{\mathcal{F}}\boldsymbol{g}\boldsymbol{\mathcal{D}}-\boldsymbol{\mathcal{D}}\boldsymbol{g}\boldsymbol{\mathcal{F}})\boldsymbol{g}(\mathbf{U}_\mathrm{a}\otimes\mathbf{U}_\mathrm{a})\nonumber\\
 &\hspace{0.4cm}+(\mathbf{U}_\mathrm{a}\otimes\mathbf{U}_\mathrm{a})\boldsymbol{g}(\boldsymbol{\mathcal{D}}\boldsymbol{g}\boldsymbol{\mathcal{F}}-\boldsymbol{\mathcal{F}}\boldsymbol{g}\boldsymbol{\mathcal{D}})],
 \label{eq:Topt}
\end{align}
\begin{equation}
 \mathbf{T}_\mathrm{ost}=(p_\mathrm{oes}+p_\mathrm{oms})\Big(\frac{\mathbf{U}_\mathrm{a}\otimes\mathbf{U}_\mathrm{a}}{c^2}-\boldsymbol{g}\Big),
 \label{eq:Tost}
\end{equation}
\begin{equation}
 \mathbf{T}_\mathrm{mech}=p_\mathrm{mech}\Big(\frac{\mathbf{U}_\mathrm{a}\otimes\mathbf{U}_\mathrm{a}}{c^2}-\boldsymbol{g}\Big).
 \label{eq:Tmech}
\end{equation}
Here $\mathbf{U}_\mathrm{a}=\gamma_{\mathbf{v}_\mathrm{a}}(c,\mathbf{v}_\mathrm{a})$ is the four-velocity of the material, and the contravariant forms of the electromagnetic field tensor $\boldsymbol{\mathcal{F}}$ and the electromagnetic displacement tensor $\boldsymbol{\mathcal{D}}$ are given by
\begin{equation}
\!\!\boldsymbol{\mathcal{F}}=\left[\begin{array}{cccc}
0 & -E_x/c & -E_y/c & -E_z/c\\
E_x/c & 0 & -B_z & B_y\\
E_y/c & B_z & 0 & -B_x\\
E_z/c & -B_y & B_x & 0
\end{array}\right],
\label{eq:Ftensor}
\end{equation}
\begin{equation}
 \boldsymbol{\mathcal{D}}=\left[\begin{array}{cccc}
0 & -D_xc & -D_yc & -D_zc\\
D_xc & 0 & -H_z & H_y\\
D_yc & H_z & 0 & -H_x\\
D_zc & -H_y & H_x & 0
\end{array}\right]\!.
\label{eq:Dtensor}
\end{equation}
The electro- and magnetostrictive pressures $p_\mathrm{oes}$ and $p_\mathrm{oms}$, the mechanical pressure $p_\mathrm{mech}$, and the mass density $\rho_\mathrm{a}$ are Lorentz scalars.

The Lorentz invariant form of the electrostrictive pressure is given by $p_\mathrm{oes}=-\frac{1}{2}[\mathbf{P}\cdot\mathbf{E}-\mathbf{v}_\mathrm{a}\cdot(\mathbf{P}\times\mathbf{B})]$ and the magnetostrictive pressure is given by $p_\mathrm{oms}=-\frac{1}{2}[\mathbf{M}\cdot\mathbf{B}+\mathbf{v}_\mathrm{a}\cdot(\mathbf{M}\times\mathbf{E})/c^2]$. The second terms of these expressions are expected to be related to the R\"ontgen and Aharonov-Casher interactions of previous literature since they have similar forms \cite{Horsley2006}. As conventional, the factor $\frac{1}{2}$ is related to the fact that we are here dealing with induced dipoles and not permanent ones or a system of free charges. In the special case of the laboratory frame, where the atomic velocity is negligible as $\mathbf{v}_\mathrm{a}\approx\mathbf{0}$, we have $p_\mathrm{oes}\approx-\frac{1}{2}\mathbf{P}\cdot\mathbf{E}$ and $p_\mathrm{oms}\approx-\frac{1}{2}\mathbf{M}\cdot\mathbf{B}$ in agreement with Eqs.~\eqref{eq:poes} and \eqref{eq:poms}.

The SEM tensor of the material in Eq.~\eqref{eq:Tmat} is of the well-known form \cite{Misner1973,Dirac1996}. The optical generalized Abraham SEM tensor in Eq.~\eqref{eq:Topt} has appeared in previous literature in Refs.~\cite{Obukhov2008,Makarov2011,Partanen2021b}, and it has been generalized for dispersive materials in Ref.~\cite{Partanen2022b}. The SEM tensors of the optostriction and mechanical pressure in Eqs.~\eqref{eq:Tost} and \eqref{eq:Tmech} are of the well-known form of the pressure term appearing in the SEM tensor of a perfect fluid \cite{Misner1973}.

In the special case of the laboratory frame, where the atomic velocity is negligibly small, the general expressions of the SEM tensors in Eqs.~\eqref{eq:Tmat}--\eqref{eq:Tmech} reduce to the following simple formulas:
\begin{equation}
 \mathbf{T}_\mathrm{mat}
 =\left[\begin{array}{cc}
\rho_\mathrm{a}c^2 & \rho_\mathrm{a}\mathbf{v}_\mathrm{a}^Tc\\
\rho_\mathrm{a}\mathbf{v}_\mathrm{a}c & \rho_\mathrm{a}\mathbf{v}_\mathrm{a}\otimes\mathbf{v}_\mathrm{a}
\end{array}\right],
\end{equation}
\begin{align}
 &\mathbf{T}_\mathrm{owm}\nonumber\\
 &\!\!=\!
 \bigg[\begin{array}{cc}
  \!\!\frac{1}{2}(\mathbf{E}\!\cdot\!\mathbf{D}\!+\!\mathbf{H}\!\cdot\!\mathbf{B}) & \frac{1}{c}(\mathbf{E}\!\times\!\mathbf{H})^T\\
  \frac{1}{c}\mathbf{E}\!\times\!\mathbf{H} & \frac{1}{2}(\mathbf{E}\!\cdot\!\mathbf{D}\!+\!\mathbf{H}\!\cdot\!\mathbf{B})\mathbf{I}-\mathbf{E}\!\otimes\!\mathbf{D}-\mathbf{H}\!\otimes\!\mathbf{B}\!\!
 \end{array}\bigg],
\end{align}
\begin{equation}
 \mathbf{T}_\mathrm{ost}
 =\left[\begin{array}{cc}
0 & \mathbf{0}\\
\mathbf{0} & (p_\mathrm{oes}+p_\mathrm{oms})\mathbf{I}
\end{array}\right],
\end{equation}
\begin{equation}
 \mathbf{T}_\mathrm{mech}
 =\left[\begin{array}{cc}
0 & \mathbf{0}\\
\mathbf{0} & p_\mathrm{mech}\mathbf{I}
\end{array}\right].
\end{equation}
Here, for the optoelectro- and optomagnetostrictive pressures $p_\mathrm{oes}$ and $p_\mathrm{oms}$, one can use the laboratory frame expressions, given in Eqs.~\eqref{eq:poes} and \eqref{eq:poms}.

\section{\label{sec:discussion}Discussion}

An obvious question arising from the present theory of optostriction is how it influences existing theories of the forces of light \cite{Anghinoni2022}. We discuss in particular the implications to the previous mass-polariton theory \cite{Partanen2017c,Partanen2019a,Partanen2019b,Partanen2021b,Partanen2022b,Partanen2018a,Partanen2018b,Partanen2022a}. As we pointed out in the analysis of Fig.~\ref{fig:curves}, the optostrictive force density does not lead to net momentum transfer between the field and the material. Thus, it does not contribute to the transfer of wave momentum by the optical field or the net momentum carried by the material. Therefore, the optostrictive force density does not either lead to net mass transfer of the material, i.e., it does not contribute to the transferred mass of the mass-polariton state. Accordingly, the law of constant velocity of the center of energy of an isolated system remains fulfilled within this generalization of the mass-polariton theory.

The theory can, however, lead to a locally nonzero momentum density of the material, and actually, in homogeneous materials, the optostrictive force density for light beams and pulses typically \emph{dominates} over the optical wave momentum force density. In particular, the optostrictive force density has a radial component, which for typical light pulses and beams, is larger than longitudinal force density component of the Abraham force term, which contributes to the transfer of wave momentum. Consequently, the optostrictive force density must be included in the force densities of light beams and pulses, and it is especially important in the modeling of experimentally observed displacements of the material caused by optical fields.

Both the optical wave momentum force density $\mathbf{f}_\mathrm{owm}$ and the optostrictive force density $\mathbf{f}_\mathrm{ost}$ act in the elastic wave equation of the material as driving forces of the dynamics. For short optical pulses, in the regime, where the field is nonzero, $\mathbf{f}_\mathrm{owm}$ and $\mathbf{f}_\mathrm{ost}$ dominate over reactive elastic forces. This means that, in the time and displacement scale of the optical pulse, the changes in the interatomic distances are so small that elastic forces have negligible contribution. Thus, the optical pulse gives a forced displacement and impulse of atoms that can be calculated from the sum of $\mathbf{f}_\mathrm{owm}$ and $\mathbf{f}_\mathrm{ost}$. After the field fades out, these displacements and related momenta form the initial state of the elastic relaxation, which is seen as sound waves. In this respect, $\mathbf{f}_\mathrm{owm}$ and $\mathbf{f}_\mathrm{ost}$ behave qualitatively in the same way.

From the point of view of the SEM tensors, discussed in Sec.~\ref{sec:tensors}, the relativistic covariance of the theory is preserved since the optostriction is described by the SEM tensor $\mathbf{T}_\mathrm{ost}$ in Eq.~\eqref{eq:Tost}, whose relativistic covariance is independent of the covariance of $\mathbf{T}_\mathrm{owm}$ in Eq.~\eqref{eq:Topt}. The relativistic covariance of the theory would be preserved even if the dissipation terms introduced in the calculation of the optostrictive force density were neglected. However, in this case, the form of the optostrictive pressure in a general inertial frame would become more complicated.

We can conclude that the optostrictive force density is an essential addition to the optical wave momentum force density of the mass-polariton theory used in previous works. It is important in describing the position- and time-dependent dynamics of the material, but it does not influence the relativistic covariance of the theory. The theory of optoelectro- and optomagnetostrictive force densities presented in this work gives a possibility to calculate the full position and time-dependent dynamics of the material under the influence of the optical field. Thus, time or harmonic cycle averaging is not needed within this theory.

From the experimental point of view, it is interesting to compare the optoelectro- and optomagnetostrictive pressures $p_\mathrm{oes}$ and $p_\mathrm{oms}$ to the electrostrictive and magnetostrictive pressures $p_\mathrm{es}$ and $p_\mathrm{ms}$, derived originally for stationary fields \cite{Brevik2018a}. It is found that these pressures are related by $p_\mathrm{es}=\frac{\varepsilon_\mathrm{r}+2}{3}p_\mathrm{oes}$ and $p_\mathrm{ms}=\frac{\mu_\mathrm{r}+2}{3\mu_\mathrm{r}}p_\mathrm{oms}$. While the electrostrictive pressure $p_\mathrm{es}$ explains the results of the classic Hakim-Higham experiment \cite{Hakim1962} for stationary fields, the optoelectrostrictive pressure $p_\mathrm{oes}$ explains the results of the recent measurement by Astrath \emph{et al.}~\cite{Astrath2022} for an optical field. At optical frequencies, in the case of water, $p_\mathrm{oes}$ is about 20\% smaller than $p_\mathrm{es}$, so the relative accuracy of the measurement does not need to be exceptionally high to observe which one of the formulas agrees with the results better. However, since the amount of quantitatively accurate measurements of electrostriction is very limited, it is desired to verify the reproducibility of the results, to narrow down error margins, and to carry out similar measurements for other materials.

The condition of a linear, isotropic material and the applicability of the Clausius-Mossotti relation limit the possibility to use the present theory for all materials. However, extension of the present theory to cover materials with different relations between the polarizability of the material and the macroscopic electric field can be developed following the principles presented in this work.

Regarding the anisotropy of the electrostrictive tensor of a deformed material, we note that the anisotropy is negligible for typical field strengths in common photonic materials. In the case of light pulses, the optostrictive forces give an impulse to the material atoms. The associated momenta later result in atomic displacement. These displacements develop at acoustic velocities and do not have time to become significant during the short optical transient. Anisotropies resulting from faster processes, such as the Kerr effect, could in principle take place. However, these effects are also negligible for common photonic materials. For example, in a recent work in Ref.~\cite{Astrath2023}, it was found that the change generated \emph{in the refractive index} of water due to the Kerr effect is negligible being of the order of 10$^{-10}$. Furthermore, the anisotropy in the electrostrictive tensor must vanish in the limit of low field strength.

\section{\label{sec:conclusions}Conclusions}

In conclusion, we have derived a time-dependent theory of force densities generated on the material by an optical field. The theory extends the previous mass-polariton theory of light to include optoelectro- and optomagnetostrictive force densities, which arise from the atomic density dependence of the energy density of the electric and magnetic fields. By introducing additional dissipation terms, nonexistent in the conventional theory of electrostriction and magnetostriction, we were able to explain the difference between the existing experimental results for a stationary field and for an optical field. The dissipation terms, we have introduced, are necessary to conserve the total energy when the optostrictive force does contraction work on the material. In the present work, we have determined the magnitude of this dissipation starting from the Lorentz force model of the optostrictive force and the requirement of the conservation of energy during the contraction of the material. Developing a physical model for this dissipation is a topic of further work. We have shown that the theory is relativistically covariant, meaning that it can be applied to an arbitrary inertial observer independent of its velocity with respect to the material. This is a strong condition limiting the number of possible theories. The understanding of electrostrictive and magnetostrictive forces at optical frequencies, developed in the present theory, is expected to revive interest in experimental studies of optical forces in various photonic materials. We also expect that the unified electromagnetic force theory for optical fields will open new approaches to develop our understanding of physics and engineering of thermal and acousto-optical coupling of light and dielectrics.

\begin{acknowledgments}
This work has been funded by the Academy of Finland under Contract No.~318197 and 349971. B.A. and N.G.C.A acknowledge CNPq (409403/2018-0, 304738/2019-0) and CAPES (Finance Code 001) for financial support.
\end{acknowledgments}

\appendix

\section{\label{apx:fem}Lorentz force density}

In this appendix, we present the total electromagnetic force density $\mathbf{f}_\mathrm{em}$ based on the Lorentz force law. This force densty can be split into the force density $\mathbf{f}_\mathrm{e}$ for induced electric dipoles and $\mathbf{f}_\mathrm{m}$ for induced magnetic dipoles. These parts and their sum are given in the subsections below.

\subsection{\label{apx:fe}Force density on induced electric dipoles}

For electric dipoles generating the polarization field $\mathbf{P}$, the Lorentz force density is well known to be given by \cite{Gordon1973,Hinds2009,Barnett2006,Stenholm1986,Penfield1967,Landau1984}
\begin{align}
 \mathbf{f}_\mathrm{e} &=(\mathbf{P}\cdot\boldsymbol{\nabla})\mathbf{E}+\frac{\partial\mathbf{P}}{\partial t}\times\mathbf{B}.
 \label{eq:fe}
\end{align}
Using $\mathbf{P}=\varepsilon_0(\varepsilon_\mathrm{r}-1)\mathbf{E}$ and applying the mathematical identity
$(\mathbf{E}\cdot\boldsymbol{\nabla})\mathbf{E}=\nabla(\frac{1}{2}|\mathbf{E}|^2)-\mathbf{E}\times(\boldsymbol{\nabla}\times\mathbf{E})$
with Faraday's law $\boldsymbol{\nabla}\times\mathbf{E}=-\partial\mathbf{B}/\partial t$ and the product rule of differentiation, we can rewrite the force density in Eq.~\eqref{eq:fe} as
\begin{equation}
 \mathbf{f}_\mathrm{e}=\frac{1}{2}\nabla(\mathbf{P}\cdot\mathbf{E})-\frac{1}{2}\varepsilon_0|\mathbf{E}|^2\nabla\varepsilon_\mathrm{r}+\frac{\partial}{\partial t}(\mathbf{P}\times\mathbf{B}).
 \label{eq:fe2}
\end{equation}
The derivation of the force density $\mathbf{f}_\mathrm{e}$ from the Lorentz forces on the individual charges of the induced electric dipole is presented briefly below.

The force applied on a single electric charge $\pm q_\mathrm{e}$ in an electromagnetic field at position $\mathbf{r}_{\pm q_\mathrm{e}}$ is known as the Lorentz force \cite{Jackson1999,Landau1989}, and it is given by
\begin{equation}
 \mathbf{F}_{\pm q_\mathrm{e}}(t)=\pm q_\mathrm{e}\Big[\mathbf{E}(\mathbf{r}_{\pm q_\mathrm{e}},t)+\frac{d\mathbf{r}_{\pm q_\mathrm{e}}}{dt}\times\mathbf{B}(\mathbf{r}_{\pm q_\mathrm{e}},t)\Big].
\end{equation}
Let us take a point $\mathbf{r}_0$ as the center of mass of two charges, which make the electric dipole. Then, we write the macroscopic electric and magnetic fields, $\mathbf{E}$ and $\mathbf{B}$, around $\mathbf{r}_0$ by using two first terms of their truncated Taylor series as
\begin{equation}
 \mathbf{E}(\mathbf{r},t)=\mathbf{E}(\mathbf{r}_0,t)
 +[(\mathbf{r}-\mathbf{r}_0)\cdot\boldsymbol{\nabla}]\mathbf{E}(\mathbf{r},t)|_{\mathbf{r}=\mathbf{r}_0}.
 \label{eq:TaylorE}
\end{equation}
\begin{equation}
 \mathbf{B}(\mathbf{r},t) =\mathbf{B}(\mathbf{r}_0,t)
 +[(\mathbf{r}-\mathbf{r}_0)\cdot\boldsymbol{\nabla}]\mathbf{B}(\mathbf{r},t)|_{\mathbf{r}=\mathbf{r}_0}.
 \label{eq:TaylorB}
\end{equation}

Next, we write the net force on the center of mass of an electric dipole made of two opposite charges, given by $\mathbf{F}_\mathrm{e}=\mathbf{F}_{q_\mathrm{e}}+\mathbf{F}_{-q_\mathrm{e}}$. Using the field approximations in Eqs.~\eqref{eq:TaylorE} and \eqref{eq:TaylorB} with dropping out the second term of Eq.~\eqref{eq:TaylorB}, since the resulting force density terms are negligible in comparison with the terms included \cite{Stenholm1986}, the net force on an electric dipole then becomes \cite{Gordon1973,Hinds2009,Barnett2006,Stenholm1986,Penfield1967,Landau1984}
\begin{align}
 \mathbf{F}_\mathrm{e}(t)
 &=q_\mathrm{e}\Big[\mathbf{E}(\mathbf{r}_q,t)-\mathbf{E}(\mathbf{r}_{-q},t)\nonumber\\
 &\hspace{0.5cm}+\frac{d\mathbf{r}_q}{dt}\times\mathbf{B}(\mathbf{r}_q,t)-\frac{d\mathbf{r}_{-q}}{dt}\times\mathbf{B}(\mathbf{r}_{-q},t)\Big]\nonumber\\
 &=(\mathbf{p}\cdot\boldsymbol{\nabla})\mathbf{E}(\mathbf{r},t)|_{\mathbf{r}=\mathbf{r}_0}
 +\frac{d\mathbf{p}}{dt}\times\mathbf{B}(\mathbf{r}_0,t).
 \label{eq:Fd}
\end{align}
Here we have defined the electric dipole moment as $\mathbf{p}=q_\mathrm{e}(\mathbf{r}_q-\mathbf{r}_{-q})$. By defining the polarization field $\mathbf{P}$ as the dipole moment density through $\mathbf{P}=n_\mathrm{a}\mathbf{p}$, where $n_\mathrm{a}$ is the number density of electric dipoles, the force density $\mathbf{f}_\mathrm{e}=n_\mathrm{a}\mathbf{F}_\mathrm{e}$ is then obtained as given in Eq.~\eqref{eq:fe}.

\subsection{Force density on induced magnetic dipoles}

For magnetic dipoles, the derivation of the Lorentz force density is less straightforward than for electric dipoles and involves the enigma of the hidden momentum \cite{Griffiths2015,Correa2020}.  The resulting force density is given by \cite{Astrath2022,Anghinoni2023}
\begin{equation}
 \mathbf{f}_\mathrm{m}=(\mathbf{M}\cdot\boldsymbol{\nabla})\mathbf{B}+\mathbf{M}\times(\boldsymbol{\nabla}\times\mathbf{B})-\frac{1}{c^2}\frac{\partial}{\partial t}(\mathbf{M}\times\mathbf{E}).
 \label{eq:fm}
\end{equation}
Using the constitutive relations in Eq.~\eqref{eq:B}, which give $\mathbf{M}=\frac{\mu_\mathrm{r}-1}{\mu_0\mu_\mathrm{r}}\mathbf{B}$, and applying the mathematical identity
$(\mathbf{B}\cdot\boldsymbol{\nabla})\mathbf{B}=\nabla(\frac{1}{2}|\mathbf{B}|^2)-\mathbf{B}\times(\boldsymbol{\nabla}\times\mathbf{B})$ and the product rule of differentiation, we can rewrite the force density in Eq.~\eqref{eq:fm} as
\begin{equation}
 \mathbf{f}_\mathrm{m}=\frac{1}{2}\nabla(\mathbf{M}\cdot\mathbf{B})-\frac{1}{2}\mu_0|\mathbf{H}|^2\nabla\mu_\mathrm{r}-\frac{1}{c^2}\frac{\partial}{\partial t}(\mathbf{M}\times\mathbf{E}).
 \label{eq:fm2}
\end{equation}

\subsection{Total force density on atomic dipoles}

The total electromagnetic force density $\mathbf{f}_\mathrm{em}$ in Eq.~\eqref{eq:fem} on a material made of both induced electric and magnetic dipoles can be written as a sum of the electric and magnetic parts in Eqs.~\eqref{eq:fe} and \eqref{eq:fm} as $\mathbf{f}_\mathrm{em}=\mathbf{f}_\mathrm{e}+\mathbf{f}_\mathrm{m}$, resulting in
\begin{align}
 \mathbf{f}_\mathrm{em} &=(\mathbf{P}\cdot\boldsymbol{\nabla})\mathbf{E}+(\mathbf{M}\cdot\boldsymbol{\nabla})\mathbf{B}+\mathbf{M}\times(\boldsymbol{\nabla}\times\mathbf{B})\nonumber\\
 &\hspace{0.5cm}+\frac{\partial\mathbf{P}}{\partial t}\times\mathbf{B}-\frac{1}{c^2}\frac{\partial}{\partial t}(\mathbf{M}\times\mathbf{E}).
 \label{eq:fLorentz}
\end{align}
Alternatively, using Eqs.~\eqref{eq:fe2} and \eqref{eq:fm2} for $\mathbf{f}_\mathrm{e}$ and $\mathbf{f}_\mathrm{m}$ and applying the constitutive relations in Eqs.~\eqref{eq:D} and \eqref{eq:B} to combine the time derivative terms of $\mathbf{f}_\mathrm{e}$ and $\mathbf{f}_\mathrm{m}$ into a single term $\frac{n^2-1}{c^2}\frac{\partial}{\partial t}(\mathbf{E}\times\mathbf{H})$, where we have used $\varepsilon_\mathrm{r}\mu_\mathrm{r}=n^2$ and taken this factor out of the time derivative, we obtain
\begin{align}
 \mathbf{f}_\mathrm{em}&=\frac{1}{2}\nabla(\mathbf{P}\cdot\mathbf{E})+\frac{1}{2}\nabla(\mathbf{M}\cdot\mathbf{B})-\frac{1}{2}\varepsilon_0|\mathbf{E}|^2\nabla\varepsilon_\mathrm{r}\nonumber\\
 &\hspace{0.5cm}-\frac{1}{2}\mu_0|\mathbf{H}|^2\nabla\mu_\mathrm{r}+\frac{n^2-1}{c^2}\frac{\partial}{\partial t}(\mathbf{E}\times\mathbf{H}).
 \label{eq:Lorentz2}
\end{align}

\section{\label{apx:stationary}Electrostriction and magnetostriction for stationary fields}

In the calculation of the conventional electro- and magnetostrictive force densities for stationary fields, one makes use of the conservation of energy when a small change is made in the density of the material \cite{Landau1984,Boyd2008}. The Clausius-Mossotti relation in Eq.~\eqref{eq:CMe} and its magnetic analog in Eq.~\eqref{eq:CMm} show that the increase of the density of the material corresponds to the increase of the relative permittivity and permeability. This can be concluded from the derivatives of the relative permittivity and permeability with respect to the atomic density, for which Eqs.~\eqref{eq:CMe} and \eqref{eq:CMm} give $n_\mathrm{a}\frac{\partial\varepsilon_\mathrm{r}}{\partial n_\mathrm{a}}=\frac{1}{3}(\varepsilon_\mathrm{r}-1)(\varepsilon_\mathrm{r}+2)$ and $n_\mathrm{a}\frac{\partial\mu_\mathrm{r}}{\partial n_\mathrm{a}}=\frac{1}{3}(\mu_\mathrm{r}-1)(\mu_\mathrm{r}+2)$. Thus, if the fields $\mathbf{D}$ and $\mathbf{B}$ are kept constant, the electric and magnetic field energy densities $W_\mathrm{e}=\frac{1}{2\varepsilon_0\varepsilon_\mathrm{r}}|\mathbf{D}|^2$ and $W_\mathrm{m}=\frac{1}{2\mu_0\mu_\mathrm{r}}|\mathbf{B}|^2$ decrease when the density of the material is increased. Then, following the principle of virtual work, the field energy density acts a potential energy, which tends to compress the material. The reduction of the field energy density must be equal to the increase of the elastic and thermal energy densities of the material. At equilibrium, the electro- and magnetostrictive force density is then equal to the thermodynamical force density corresponding to the compressibility of the material for the given thermodynamical process \cite{Boyd2008}. For an isothermal process taking place at constant temperature $T$, we then obtain the electrostrictive force density $\mathbf{f}_\mathrm{es}$ and magnetostrictive force density $\mathbf{f}_\mathrm{ms}$ from the electric and magnetic free energy densities $F_\mathrm{e}=W_\mathrm{e}$ and $F_\mathrm{m}=W_\mathrm{m}$ as
\begin{align}
 \mathbf{f}_\mathrm{es} &=-\nabla\Big[n_\mathrm{a}\Big(\frac{\partial F_\mathrm{e}}{\partial n_\mathrm{a}}\Big)_{\mathbf{D},T}\Big]\nonumber\\
 &=\frac{1}{2}\varepsilon_0\nabla\Big[n_\mathrm{a}\Big(\frac{\partial\varepsilon_\mathrm{r}}{\partial n_\mathrm{a}}\Big)_T|\mathbf{E}|^2\Big]\nonumber\\
 &=\frac{1}{2}\nabla(\mathbf{P}\cdot\mathbf{E}_\mathrm{eff}),
 \label{eq:fes}
\end{align}
\begin{align}
 \mathbf{f}_\mathrm{ms} &=-\nabla\Big[n_\mathrm{a}\Big(\frac{\partial F_\mathrm{m}}{\partial n_\mathrm{a}}\Big)_{\mathbf{B},T}\Big]\nonumber\\
 &=\frac{1}{2}\mu_0\nabla\Big[n_\mathrm{a}\Big(\frac{\partial\mu_\mathrm{r}}{\partial n_\mathrm{a}}\Big)_T|\mathbf{H}|^2\Big]\nonumber\\
 &=\frac{1}{2}\nabla(\mathbf{M}\cdot\mathbf{B}_\mathrm{eff}).
 \label{eq:fms}
\end{align}
Equations \eqref{eq:fes} and \eqref{eq:fms} define the conventional stationary electrostrictive and magnetostrictive pressures $p_\mathrm{es}$ and $p_\mathrm{ms}$ through $\mathbf{f}_\mathrm{es}=-\nabla p_\mathrm{es}$ and $\mathbf{f}_\mathrm{ms}=-\nabla p_\mathrm{ms}$. Thus, $p_\mathrm{es}=-\frac{1}{2}\varepsilon_0n_\mathrm{a}\big(\frac{\partial\varepsilon_\mathrm{r}}{\partial n_\mathrm{a}}\big)_T|\mathbf{E}|^2=-\frac{1}{6}\varepsilon_0(\varepsilon_\mathrm{r}-1)(\varepsilon_\mathrm{r}+2)|\mathbf{E}|^2$ and $p_\mathrm{ms}=-\frac{1}{2}\mu_0n_\mathrm{a}\big(\frac{\partial\mu_\mathrm{r}}{\partial n_\mathrm{a}}\big)_T|\mathbf{H}|^2=-\frac{1}{6}\mu_0(\mu_\mathrm{r}-1)(\mu_\mathrm{r}+2)|\mathbf{H}|^2$ \cite{Boyd2008,Landau1984}. The stationary electrostrictive pressure has been quantitatively measured for selected non-polar isotropic liquid dielectrics in the classic Hakim-Higham experiment \cite{Hakim1962}.

Alternatively, if the fields $\mathbf{E}$ and $\mathbf{H}$ are kept constant, the electric and magnetic field energy densities $W_\mathrm{e}=\frac{1}{2}\varepsilon_0\varepsilon_\mathrm{r}|\mathbf{E}|^2$ and $W_\mathrm{m}=\frac{1}{2}\mu_0\mu_\mathrm{r}|\mathbf{H}|^2$ increase when the density of the material is increased. At first sight, this seems to be in contradiction with the argumentation above. However, in this case one must do external work to preserve the values of the fields $\mathbf{E}$ and $\mathbf{H}$ at the same time when the density of the material is increased. When this external work is accounted for, the free energy densities $F_\mathrm{e}$ and $F_\mathrm{m}$ become replaced by $\tilde F_\mathrm{e}=-W_\mathrm{e}$ and $\tilde F_\mathrm{m}=-W_\mathrm{m}$, and we obtain $\mathbf{f}_\mathrm{es}=-\nabla\big[n_\mathrm{a}\big(\frac{\partial\tilde F_\mathrm{e}}{\partial n_\mathrm{a}}\big)_{\mathbf{E},T}\big]$ and $\mathbf{f}_\mathrm{ms}=-\nabla\big[n_\mathrm{a}\big(\frac{\partial\tilde F_\mathrm{m}}{\partial n_\mathrm{a}}\big)_{\mathbf{H},T}\big]$. As a result, the values of the electrostrictive and magnetostrictive force densities are not changed from those obtained by using Eqs.~\eqref{eq:fes} and \eqref{eq:fms}. In an example of a dielectric material placed between capacitor plates, keeping the field $\mathbf{D}$ constant corresponds to having fixed charges on the capacitor plates, while keeping the field $\mathbf{E}$ constant corresponds to having fixed potential between the capacitor plates (Sec.~4.7 of Ref.~\cite{Jackson1999}). In the latter case, one must do work to increase the charges on the capacitor plates when the permittivity of the dielectric is increased.

Note that the stationary electrostrictive and magnetostrictive force densities in Eqs.~\eqref{eq:fes} and \eqref{eq:fms} could also be calculated as gradient force densities from the potential energy density resulting from the Stark [$\Delta E_\mathrm{e}^{(n)}=-\frac{1}{2}(\alpha_\mathrm{e})_{ik}^{(n)}(E_\mathrm{eff})_i(E_\mathrm{eff})_k$] and Zeeman [$\Delta E_\mathrm{m}^{(n)}=-\frac{1}{2}(\alpha_\mathrm{m})_{ik}^{(n)}(B_\mathrm{eff})_i(B_\mathrm{eff})_k$] shifts \cite{Landau1977} of atomic energy levels in an external electromagnetic field. The equivalence of this approach with the last rows of Eqs.~\eqref{eq:fes} and \eqref{eq:fms} can be seen by assuming a single ground-state energy level, assuming that its polarizability and magnetizability tensors $(\alpha_\mathrm{e})_{ik}^{(n)}$ and $(\alpha_\mathrm{m})_{ik}^{(n)}$ are diagonal as $(\alpha_\mathrm{e})_{ik}^{(n)}=\alpha_\mathrm{e}\delta_{ik}$ and $(\alpha_\mathrm{m})_{ik}^{(n)}=\alpha_\mathrm{m}\delta_{ik}$, and noting that $p_i=\alpha_\mathrm{e}(E_\mathrm{eff})_i$ and $m_i=\alpha_\mathrm{m}(B_\mathrm{eff})_i$ are the electric and magnetic dipole moments of the atom.

\end{document}